\documentclass[10pt,letterpaper]{article}
\usepackage[top=0.85in,left=2.75in,footskip=0.75in]{geometry}

\usepackage{amsmath,amssymb}

\usepackage{changepage}

\usepackage{textcomp,marvosym}

\usepackage[numbers]{natbib}

\usepackage{nameref,hyperref}
\hypersetup{colorlinks=true,citecolor=red}

\usepackage{subcaption}
\usepackage{comment}

\usepackage[table]{xcolor}

\usepackage{array}

\newcolumntype{+}{!{\vrule width 2pt}}

\newlength\savedwidth

\raggedright
\usepackage[aboveskip=1pt,labelfont=bf,labelsep=period,justification=raggedright,singlelinecheck=off]{caption}

\makeatletter
\renewcommand{\@biblabel}[1]{\quad#1.}
\makeatother

\usepackage{lastpage,fancyhdr,graphicx}
\usepackage{epstopdf}
\fancyheadoffset[L]{2.25in}
\fancyfootoffset[L]{2.25in}
\usepackage[title,page]{appendix}
\usepackage{xargs}
\usepackage[colorinlistoftodos,prependcaption,textsize=tiny]{todonotes}
\reversemarginpar
\definecolor{olive}{rgb}{0.5, 0.5, 0.0}
\definecolor{Plum}{rgb}{0.56, 0.27, 0.52}
\newcommandx{\unsure}[2][1=]{\todo[linecolor=red,backgroundcolor=red!25,bordercolor=red,#1]{#2}}
\newcommandx{\change}[2][1=]{\todo[linecolor=blue,backgroundcolor=blue!25,bordercolor=blue,#1]{#2}}
\newcommandx{\info}[2][1=]{\todo[linecolor=olive,backgroundcolor=olive!25,bordercolor=olive,#1]{#2}}
\newcommandx{\improvement}[2][1=]{\todo[linecolor=Plum,backgroundcolor=Plum!25,bordercolor=Plum,#1]{#2}}
\newcommandx{\thiswillnotshow}[2][1=]{\todo[disable,#1]{#2}}

\newcommand{\incfig}{\centering\includegraphics}

\newcommand{\ie}{i.e.,}

\begin{document}
\vspace*{0.2in}

\begin{flushleft}
{\Large
\textbf\newline{Novel ultrasonic bat deterrents based on aerodynamic whistles} 
}
\newline
\\
Zhangming Zeng\textsuperscript{1\Yinyang},
Anupam Sharma\textsuperscript{1\Yinyang*},
\\
\bigskip
\textbf{1} Department of Aerospace Engineering, Iowa State University, Ames, Iowa, USA 
\\
\bigskip

\Yinyang These authors contributed equally to this work




* Corresponding author: sharma@iastate.edu

\end{flushleft}
\section*{Abstract}
Novel ultrasonic bat deterrents based on aerodynamic whistles are investigated experimentally and numerically. The baseline deterrent, a single-whistle design inspired by \citet{beeken1969fluid}, is examined first. It consists of two resonating cavities/chambers. The whistle is ``powered'' by a regulated high-pressure air supply and the performance of the whistle is examined for a range of supply pressures. Farfield acoustic measurements in the $20$ Hz - $50$ kHz frequency range are made in an anechoic chamber. The noise measurements are supplemented with two- and three-dimensional unsteady Reynolds-averaged Navier-Stokes (uRANS) simulations to investigate the mechanisms of ultrasound generation. The Ffowcs Williams-Hawkings acoustic analogy is used with the 3-D uRANS results to predict the far field radiation. The farfield acoustic predictions are in good agreement with the measurements in the anechoic chamber. The peak frequency (fundamental) of the radiated ultrasound of the baseline deterrent is approximately $23$ kHz. Harmonics and a sub-harmonic of the fundamental tone are also observed. The numerical simulations show that the two resonating chambers of the baseline deterrent operate out-of-phase and Helmholtz resonance determines the whistling frequency over the range of supply air pressure considered.

A six-whistle deterrent targeting a broad spectral coverage in the $20-50$ kHz frequency range is designed, fabricated and tested in the anechoic chamber. Each whistle in the deterrent is obtained by geometrically scaling the baseline whistle. Measurements show that the six-whistle ultrasound deterrent generates ultrasound with six dominant peaks in the designed frequency range when the supply air pressure exceeds $5$ psig. The proposed ultrasound devices can serve as effective bat deterrents for a variety of purposes including reducing bat mortality at wind turbines.



\section*{Introduction}
\label{sec:Intro}
Growth in wind energy has been driven in part by improvements in turbine design, e.g.,~\citep{rosenberg2014novel,moghadassian2016numerical,hu2015experimental}, and by numerical methods that enable such designs, e.g.,~\citep{rosenberg2016prescribed,moghadassian2018inverse,moghadassian2020designing,chen2016modeling}. For wind energy to be a truly renewable energy resource however, this growth has to occur in an ecologically responsible manner. Bat mortality at wind farms is now recognized as a serious threat to bat population and is a serious impediment to future growth of wind energy in the world. Combined with the current environmental threats to bats (i.e., White-nose Syndrome), this added cause of mortality is pushing certain species of bats towards extinction \cite{frick2020review,friedenberg2021assessing,cheng2021scope}. Several strategies to mitigate bat mortality at wind farms are being pursued. These include: (a) operational mitigation \citep{martin2017reducing}, which involves curtailing power generation, typically at low wind speeds, and (b) deterring bats away from wind farms using a variety of strategies including ultrasonic deterrents \citep{arnett2013evaluating,romano2019evaluation,weaver2020ultrasonic}, ultraviolet lights \citep{gorresen2015dim}, and textured paints on turbine towers \citep{huzzen2019does}. A serious drawback of curtailment is the associated loss of wind energy capture. Bat deterrents can significantly reduce (potentially eliminate) the need for curtailment, and hence can serve as an alternate, more efficient approach. While results of initial testing with such deterrents look promising \citep{arnett2013evaluating}, long-term effectiveness of these deterrents remains to be conclusively proven. Combinations of different technologies, such as curtailment and ultrasonic deterrents, have also been explored \cite{good2022curtailment}.

Ultrasonic deterrence works by ``jamming'' the echolocation signals of bats. This is achieved by creating high-amplitude ultrasound which can be either broadband (white noise) or tonal in nature. \citet{schirmacher2020evaluating} verified that ultrasonic deterrents with a \textit{tonal spectrum} (as opposed to broadband) are also effective at driving bats away. By concentrating the acoustic power into a few frequencies (tones), the intensity of the radiated ultrasonic tones can be increased.
%
%

Current ultrasonic deterrents utilize electromechanical transducer-driven speakers \citep{horn2008testing} to generate ultrasound. These deterrents can be mounted on turbine nacelle and tower. One example of such a deterrent is the Bat Deterrent System (BDS) developed by NRG Systems \citep{schirmacher2020evaluating}. One unit of this system consists of six sub-arrays, each  using multiple transducers to generate tonal ultrasound at a predetermined frequency. For example, a BDS could target $20$, $26$, $32$, $38$, $44$ and $50$ kHz frequencies. A six-tone BDS was found to be effective in mitigating bat fatality \cite{schirmacher2020evaluating}. However, such electromechanical devices suffer from the following: (a) they require external power which limits the possible mounting positions of the devices to the nacelle and the tower, and (b) maintenance issues due to rain/water and lightning damage. 
Aerodynamic ultrasound deterrents have also been pursued \citep{romano2019evaluation}, where compressed air is accelerated in a nozzle and ejected as a supersonic jet. The noise produced by such a jet is in the ultrasound range and a broadband spectrum in the frequency range of $20 - 100$ kHz is generated. The source of compressed air supply for such a system has to be housed in either the turbine tower or the nacelle. 

Tonal ultrasound can be efficiently generated by using aerodynamic whistles. An aerodynamic whistle is a self-sustaining oscillator that can generate high-amplitude acoustic tones \cite{chanaud1970aerodynamic}. The dominant sound frequency of such a whistle is determined by its feedback mechanism. Based on the feedback mechanism, \citet{chanaud1970aerodynamic} classified aerodynamic whistles into three classes. In Class I whistles, the feedback is provided directly by the flow instability. In Class II whistles, the feedback is provided by the sound generator, and in Class III whistles, the feedback is provided by the resonator/sound reflector. 
%

Flow-excited resonance, or whistling, has been studied extensively to understand sound generation mechanisms. A jet of fluid passing through a confined space is an aerodynamic whistle, which is observed in nature. For example, humans can whistle by blowing or sucking air through a small opening to the mouth with the lips. This produces \textit{hole tones} with the mouth cavity acting as a Helmholtz resonator \citep{rayleightheory}. The whistle frequency can be adjusted by changing the position and the shape of the tongue and modifying the resonating mouth cavity. This has been experimentally investigated by \citet{wilson1971experiments}.

Flow over a cavity also generates noise. Depending on the conditions, cavity noise can be a Class II or a Class III whistle. \citet{gloerfelt2009cavity} identified the possible mechanisms of cavity noise to be (i) Rossiter modes \citep{rossiter1964wind}, which occur because of the acoustic feedback generated when the free shear layer over the cavity interacts with its downstream edge, (ii) Helmholtz resonance, which is due to the compressibility of the fluid in the cavity, and (iii) standing-wave resonance in the cavity (depth, longitudinal and spanwise modes). Of these, Rossiter modes are associated with Class II whistles, while Helmholtz resonance and standing-wave resonance are Class III. In some cases, Helmholtz resonance can exist simultaneously with standing wave resonance in a cavity \citep{bennett2017resonant}.

Whistles are used in air-driven sound generation devices such as the steam/tea kettle. Through experiments, \citet{henrywood2013aeroacoustics} showed that the whistling mechanism in a tea kettle changes with the flow condition. At low Reynolds numbers ($Re$), the mechanism is Helmholtz resonance, which exhibits a near-constant frequency behavior. At high $Re$, the sound generation is due to vortex shedding (constant Strouhal number) at the orifice coupled with the upstream duct (spout) modes.
In a classic Helmholtz resonator, the bottle ``neck'' is clearly defined, however, for cavities with thin walls (wall thickness $\ll$ cavity opening, such as the design considered in this work), the neck is not well defined, and an equivalent neck length should be used. \citet{kinsler1999fundamentals} and \citet{ruijgrok1993elements} present ways to estimate the Helmholtz resonance frequency of thin-wall cavities.

While aerodynamic whistles are typically used to generate sounds in the human-audible frequency range ($20$ Hz - $20$ kHz), the whistle geometry can be modified to generate ultrasound. This paves the way for developing ultrasonic deterrents using aerodynamic whistles. The advantages of ultrasonic deterrents based on aerodynamic whistles include design simplicity (no moving parts), low cost, and the potential for generating high-amplitude ultrasound with minimal power (in terms of air supply) requirements.

In this work, we investigate novel ultrasonic deterrents that can generate tonal spectra in the $20 - 50$ kHz frequency range. The novelty is in the use of aerodynamic whistles and resonance to generate high-amplitude ultrasound. The geometry of the whistles are tailored to restrict the tone generation in the desired frequency range. We first examine the acoustic performance of a baseline ultrasonic whistle design inspired by \citet{beeken1969fluid}. A systematic study is conducted to unveil the ultrasound generation mechanisms and quantify the acoustic performance of the baseline whistle. The investigations include laboratory tests in the fully anechoic chamber at Iowa State University (ISU), which are supplemented with numerical results. Numerical predictions of the radiated sound are obtained by coupling the near-field flow solution with an acoustic analogy. Next, we present a six-whistle deterrent targeting multiple tones in the $20-50$ kHz frequency range. The deterrent is fabricated and tested in the ISU anechoic chamber. The measurements show that the deterrent can readily generate the desired frequency spectra.

\section*{Ultrasound whistle: baseline design}
\label{sec:baseline}
\begin{figure}[!htb]
    \centering
    \subcaptionbox{CAD model \label{fig:Baseline_BeekenIdea}}{\incfig[width=0.38\textwidth]{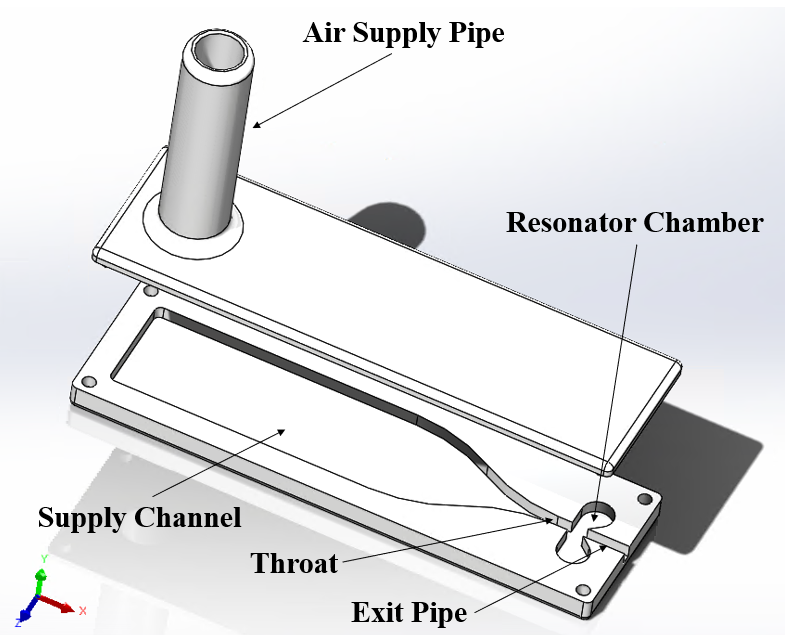}}
    \qquad
    \subcaptionbox{3D printed baseline whistle \label{fig:Baseline_3DPrint}}{\incfig[width=0.4\textwidth]{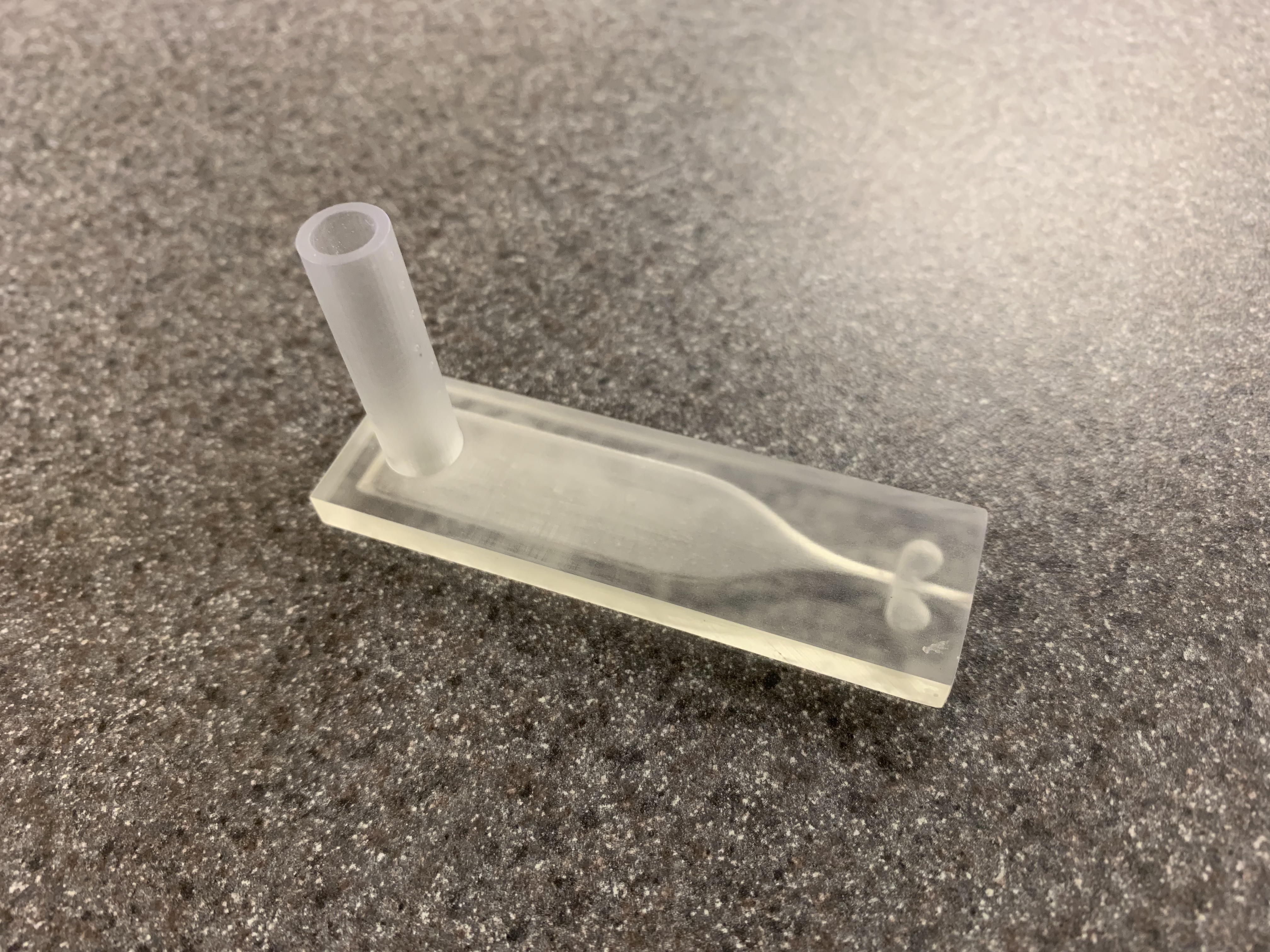}}
    \caption{The ultrasound baseline whistle design following \citet{beeken1969fluid} 
    \label{fig:Baseline_Idea_3DPrint}}
\end{figure}

Figure \ref{fig:Baseline_BeekenIdea} shows an exploded view of the baseline ultrasound whistle. The whistle consists of three parts: 1) a panel on which the flow channel is engraved, 2) a cover, and 3) a circular pipe to supply pressurized air. These parts are 3D printed using clear resin to fabricate the whistle models used in the experiments reported in this work. Pressurized air enters the whistle through the supply pipe, passes through the throat of the channel, over the orifices of the two resonating chambers, and leaves the whistle through the exit pipe. The fundamental frequency of the whistle is expected to stay constant over a wide range of supply air pressures \cite{beeken1969fluid}. By selecting proper physical dimensions the whistle can generate tones in the ultrasonic frequency range ($>20$ kHz). The working hypothesis is that this is a Class III whistle where the potential resonance mechanism is cavity noise (Helmholtz resonance, Rossiter modes, and standing waves) in the two resonating chambers.

\begin{figure}[!htb]
    \centering
    \subcaptionbox{\label{fig:Baseline_2DdesignB}}{\incfig[width=0.7\textwidth]{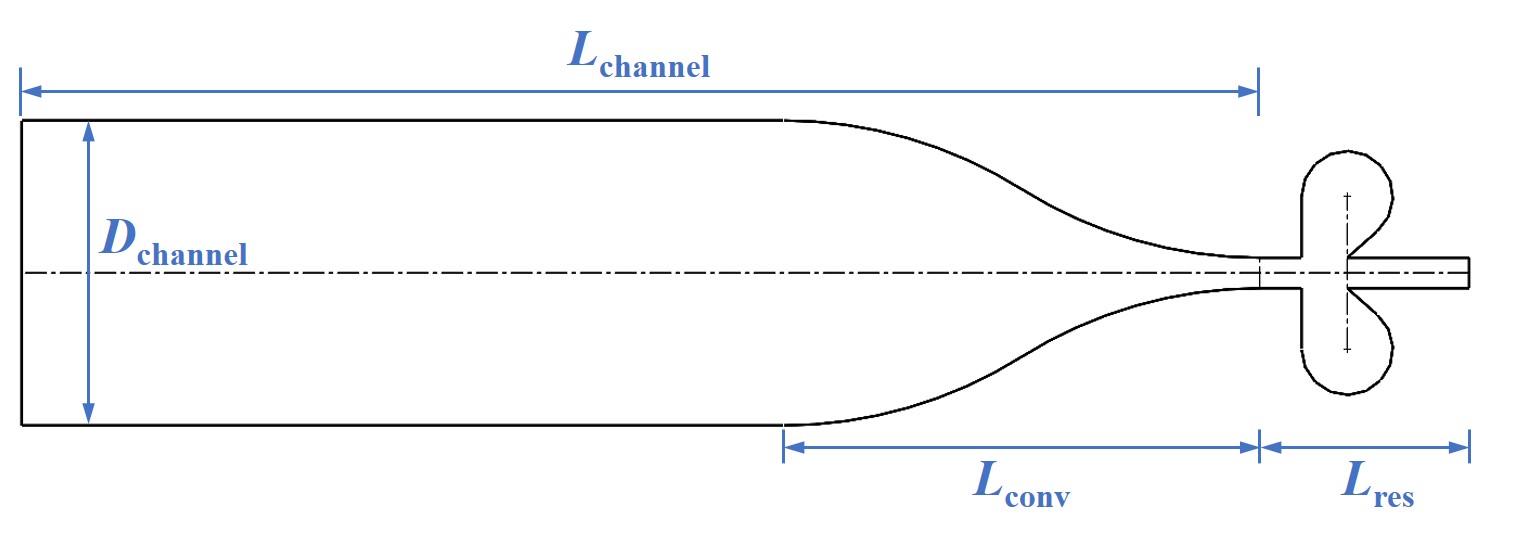}}
    \quad
    \subcaptionbox{\label{fig:Baseline_2DdesignA}}{\incfig[width=0.25\textwidth]{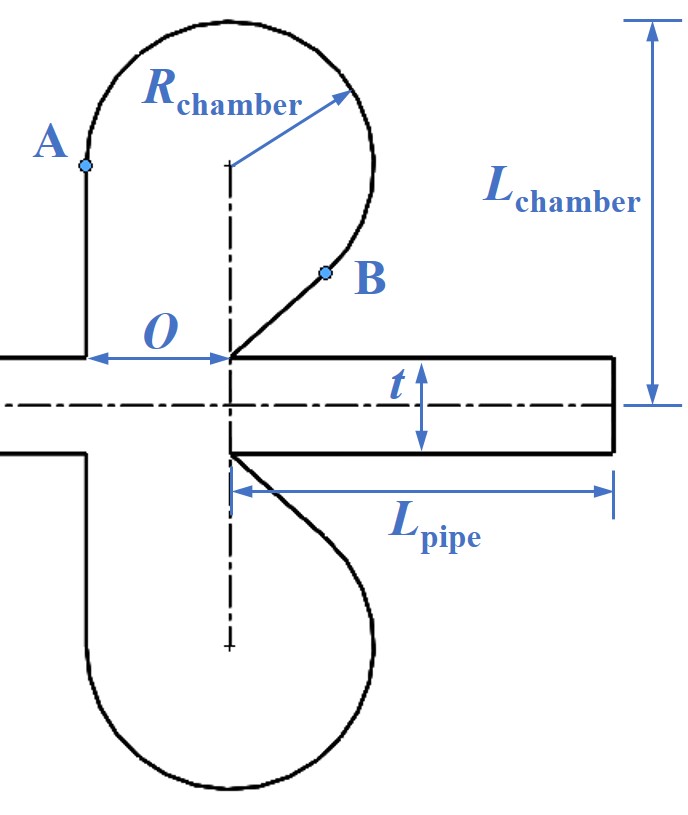}}
    \caption{Parameterization of the flow channel and resonating chambers of the baseline whistle. The dimensions are provided in Table \ref{tab:whistle_para}.\label{fig:Baseline_2Ddesign}}
\end{figure}

The acoustic performance of the baseline whistle is determined by the geometry of the flow channel. The key geometric parameters of the flow channel are identified in Fig. \ref{fig:Baseline_2Ddesign} and Table \ref{tab:whistle_para} lists their dimensional values. The whistle is sized to have a fundamental frequency between $22-25$ kHz. 

\begin{table}[!htb]
  \centering
  \caption{Key parameters of the whistle design (all lengths are in mm)\label{tab:whistle_para}}
    \begin{tabular}{c|c|c|c|c|c|c|c|c}
    $L_{\rm channel}$ & $L_{\rm conv}$ & $L_{\rm res}$ & $L_{\rm chamber}$ & $L_{\rm pipe}$ & $t$ & $O$ & $R_{\rm chamber}$ & $D_{\rm channel}$\\
    \hline
    65 & 25 & 11 & 6.4 & 6.4 & 1.6 & 2.4 & 2.4 & 2 \\
    \hline
    \end{tabular}%
\end{table}

\section*{Methods}
\label{sec:methods}
%
\subsection*{Experimental}
\label{sec:anechoic_chamber}
\begin{figure}[htb!]
    \subcaptionbox{ISU anechoic chamber\label{fig:chamber}}{\incfig[width=0.54\textwidth]{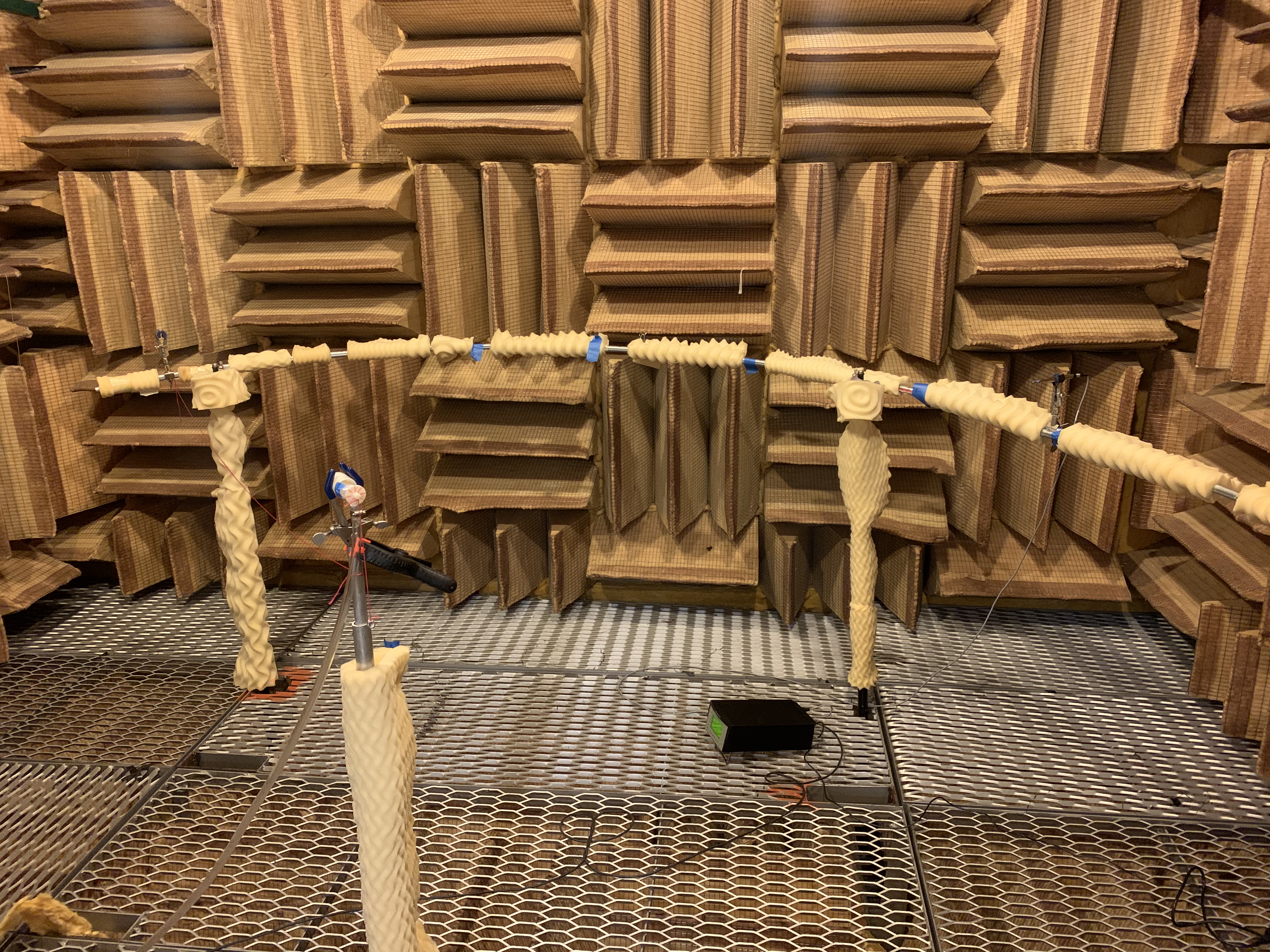}}
    \hfill
    \subcaptionbox{characterization of the chamber\label{fig:ISU_anechoic_characterization}}{\incfig[width=0.44\textwidth]{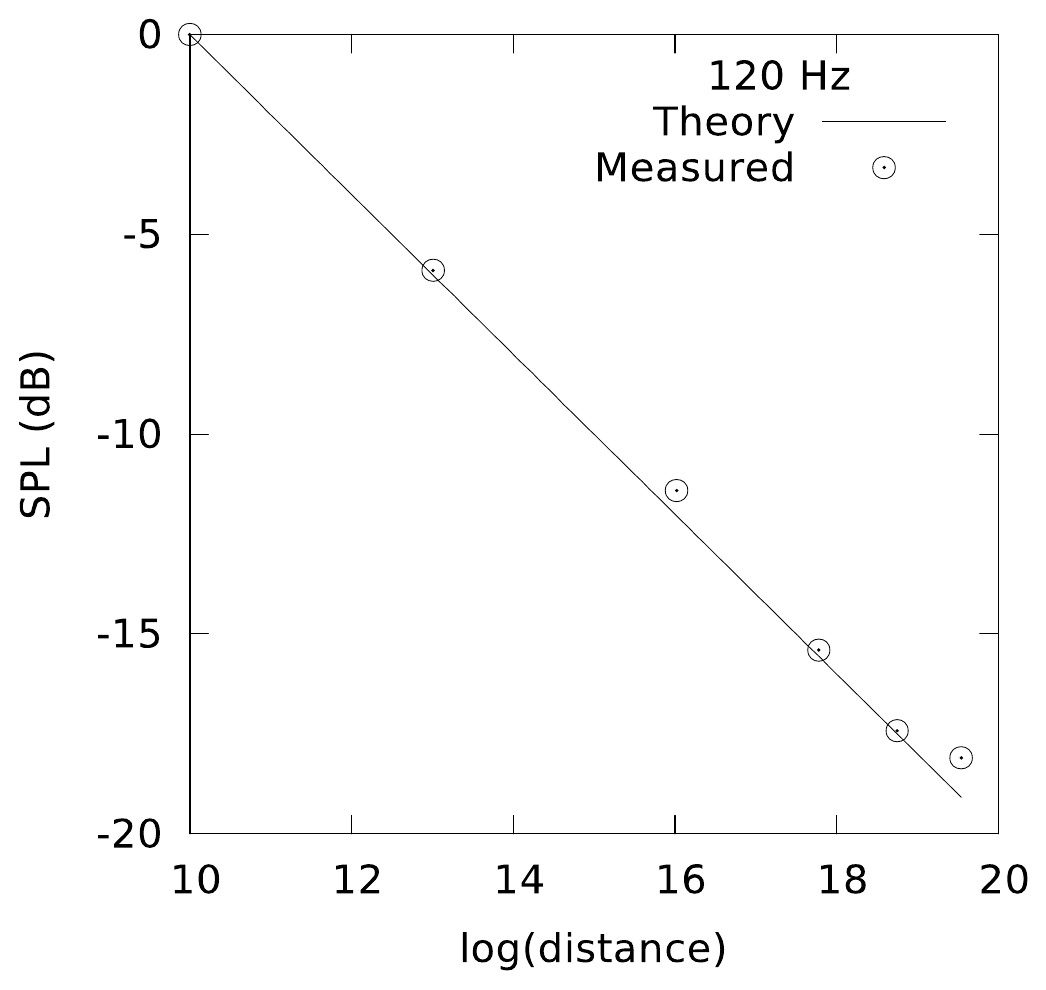}}
    \caption{The anechoic chamber facility at Iowa State University. (a) A deterrent model mounted in the anechoic chamber for acoustic testing, and (b) anechoic room soundproofing characterization, free-field radiation test result.\label{fig:anechoic_chamber}}
\end{figure}

The anechoic facility at ISU is a fully-anechoic chamber with a wire-mesh floor to support the equipment to be tested. The test section dimensions are $15'\times 15' \times 12'$ (Fig. \ref{fig:chamber}). The noise floor in the chamber is $20$ dBA and the pressure transmission coefficient is $3\times 10^{-4}$. The facility has been calibrated for free-field radiation in the chamber (see Fig. \ref{fig:ISU_anechoic_characterization}). 
%

Acoustic measurements are made using quarter-inch, free-field, B\&K 4939 microphones. The microphones are connected to a B\&K 2690 conditioning amplifier, which supports simultaneous measurements over two channels. The B\&K 2690 is connected to an NI PXIe 1073 data acquisition (DAQ) system. 
%
%
The B\&K 4939 microphone has a dynamic range up to 164 dB and a frequency range of $4$ Hz to $100$ kHz. The highest output frequency of the B\&K 2690 amplifier is $100$ kHz, thus spectral measurements up to a frequency of 50 kHz are possible. The output sensitivity of the amplifier is preset as $100$ mV/Pa.

The measured data are processed using MATLAB. Fourier analysis is performed on the time-domain pressure signal acquired by the DAQ system to obtain the power spectral density (PSD) of pressure in the frequency domain. Tonal frequencies are identified from the PSD spectra. We also report sound pressure levels (SPLs) for the tones, which are obtained by integrating the PSD spectra over frequency bands (bandwidth $2~\rm kHz$) centered at the peak frequency of the corresponding tone. The same approach is used to process the data from experiments as well as from numerical simulations. To reduce uncertainty, five independent readings are taken for each experiment in the anechoic chamber and the averaged spectra are reported.


The deterrents are driven by compressed air. The laboratory air supply pressure is around $50$ psig, which has to be reduced to lower pressures, between $0 - 10$ psig, for the deterrent devices tested. The supply air first passes through on oil/water filter for conditioning, followed by a flow control knob with a pressure gauge, and finally a high-precision volumetric flow-rate meter. For each acoustic measurement, the supply pressure and the flow rate are recorded. At the relatively low volumetric flow rates considered here ($<2$ scfm), the flow speed in the air supply system is low enough that the measured supply pressure can be approximated as the stagnation pressure driving the whistle.

The deterrent is mounted in the center of the anechoic chamber. The microphones are positioned in a semi-circle of radius $1.651$ m (see Fig. \ref{fig:chamber}) and are mounted every 15$^\circ$ along the arc to measure polar directivity (polar angle is measured from downstream of the whistle). The total radiated sound power level (PWL) from the deterrent is calculated by integrating the polar arc data over a sphere assuming no azimuthal variation. The azimuthal variation in radiated ultrasound is much less pronounced than the polar variation (see Fig.~\ref{fig:exp_directivity}).

\subsection*{Numerical}
\label{sec:numerical_methodology}
Aeroacoustics predictions are carried out using a two-step process wherein the acoustic sources are obtained in the first step via unsteady computational fluid dynamics (CFD) simulations. Ultrasound radiation to the farfield is carried out in step two using the Ffowcs Williams-Hawkings (FW-H) acoustic analogy \citep{ffowcs1969sound}. The STAR-CCM+ software is used for the CFD simulations. The numerical results supplement the radiating farfield acoustic measurements with near-field flow and acoustics information, which enables a comprehensive understanding of the flow and acoustic mechanisms involved. The simulations are performed in two and three spatial dimensions. Two-dimensional (2D) simulations are used for design studies and three-dimensional simulations for verification against measurements.

Fluid flow is governed by equations that express the conservation of mass, momentum and energy. This system of equations with an equation of state is referred to as the Navier-Stokes (N-S) equations. We solve the unsteady Reynolds-averaged Navier-Stokes (uRANS) equations, which are obtained by density-weighted (Favre) short-time averaging the Navier-Stokes equations. This averaging results in unresolved turbulent stress terms, which are modeled using a turbulence closure model. In this work, transport equations for the turbulence kinetic energy ($k$) and specific dissipation rate ($\omega$) are solved with appropriately-tuned production and dissipation terms via a $k-\omega$ turbulence model.

The density-weighted, short-time averaged N-S equations are
\begin{equation}
    \frac{\partial \Bar{\rho}}{\partial t} + \frac{\partial}{\partial x_{j}} (\Bar{\rho} \Tilde{u}_{j}) = 0,
    \label{eq:uRANS_Cont}
\end{equation}
\begin{equation}
    \frac{\partial}{\partial t} (\Bar{\rho} \Tilde{u}_{i}) + \frac{\partial}{\partial x_{j}} (\Bar{\rho} \Tilde{u}_{i} \Tilde{u}_{j})
    = -\frac{\partial \Bar{p}}{\partial x_{i}} 
    + \frac{\partial}{\partial x_{j}} (\Bar{\tau}_{i j} - \overline{\rho u^{\prime\prime}_{i} u^{\prime\prime}_{j}}),
    \label{eq:uRANS_Mom}
\end{equation}
\begin{equation}
    \frac{\partial}{\partial t}(\Bar{\rho} c_{p} \Tilde{T})
    +\frac{\partial}{\partial x_{j}}(\Bar{\rho} c_{p} \Tilde{T} \Tilde{u}_{j})
    = \frac{\partial \Bar{p}}{\partial t} + \Tilde{u}_{j} \frac{\partial \Bar{p}}{\partial x_{j}}
    + \overline{u^{\prime\prime}_{j}\frac{\partial p}{\partial x_{j}}} 
    + \frac{\partial}{\partial x_{j}}\left(
    \kappa \frac{\partial \Tilde{T}}{\partial x_{j}} + \kappa \frac{\partial \Bar{T^{\prime\prime}}}{\partial x_{j}}
    - c_{p}~\overline{\rho T^{\prime\prime} u^{\prime\prime}_{j}}
    \right)
    + \Bar{\Phi},
    \label{eq:uRANS_Egy}
\end{equation}
where the shear stress term $\overline{\tau_{i j}}$ can be written as
\begin{equation}
    \overline{\tau_{ij}} = 
    \mu \left[
    \left( \frac{\partial \Tilde{u}_i}{\partial x_j} + \frac{\partial \Tilde{u}_j}{\partial x_i}
    \right)
    -\frac{2}{3}\delta_{ij}\frac{\partial \Tilde{u}_k}{\partial u_k}
    \right]
    + \mu \left[
    \left( \frac{\partial \overline{u_i^{\prime\prime}}}{\partial x_j} + \frac{\partial \overline{u_j^{\prime\prime}}}{\partial x_i}
    \right)
    -\frac{2}{3}\delta_{ij}\frac{\partial \overline{u_k^{\prime\prime}}}{\partial u_k}
    \right],
    \label{eq:uRANS_viscous stress}
\end{equation}
and the dissipation function $\Bar{\Phi}$ can be written as
\begin{equation}
    \Bar{\Phi} = \overline{\tau_{ij}\frac{\partial u_i}{\partial x_j}}
    = \Bar{\tau}_{i j} \frac{\partial \Tilde{u}_i}{\partial x_j} 
    + \overline{\tau_{i j} \frac{\partial u_i^{\prime\prime}}{\partial x_j}}.
    \label{eq:uRANS_dissipation func}
\end{equation}
In the above, the overline $\Bar{(~~)}$ denotes short-time averaging and the tilde $\Tilde{(~~)}$ represents density-weighted time-averaging. The superscript $^{\prime\prime}$ refers to the fluctuation of the mass-averaged variables. To close the system of equations, the Reynolds stress tensor $-\overline{\rho u_i^{\prime\prime} u_j^{\prime\prime}}$ has to be modeled. In this research, the shear stress transport (SST) $k-\omega$ model \cite{menter1994two} is used.

\subsubsection*{Acoustic prediction}
\label{sec:FWH}
\begin{figure}[htb!]
    \incfig[width=0.65\textwidth]{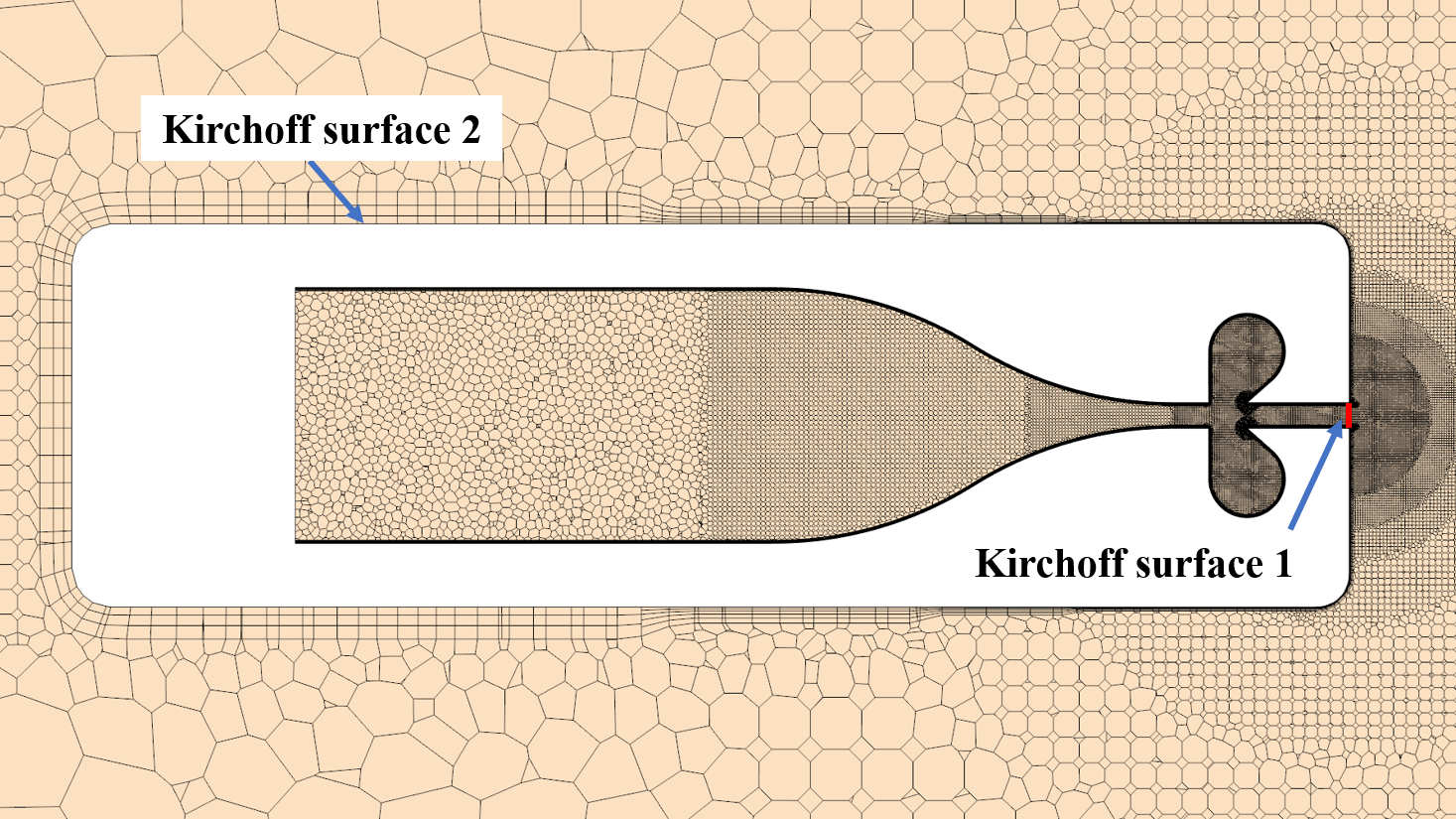} 
    \caption{The integration surface used to predict farfield ultrasound. The surface consists of a porous part (Kirchoff surface 1, shown in red) through which air can flow and an impermeable part (surface 2).\label{fig:integration_surface}}
\end{figure}

The Ffowcs Williams-Hawkings (FW-H) acoustic analogy is used to predict acoustic propagation to the farfield from the near-field sources obtained from CFD. The FW-H formulation can be expressed in the following differential form.
\begin{equation}
    \left(\frac{\partial^2}{\partial t^2} - c_o^2 \frac{\partial^2}{\partial x_i\partial x_i}\right)
    \left(H(f) \rho^\prime \right)
    =
    \frac{\partial^2}{\partial x_i\partial x_j}\left(T_{i j} H(f) \right)
    -
    \frac{\partial}{\partial x_i} \left( F_i \delta(f) \right)
    +
    \frac{\partial}{\partial t} \left(Q \delta(f) \right)
    \label{eq:FWH}
\end{equation}
where,
\begin{align}
    T_{i j } &= \rho u_i u_j + P_{i j } - c^2_o \rho^\prime \delta_{i j}, \nonumber \\
    F_i &= ( P_{i j } + \rho u_i (u_j - v_j)){\partial f}/{\partial x_j}, ~{\rm and}\\
    Q &= (\rho_o v_i + \rho (u_i - v_i)){\partial f}/{\partial x_j}. \nonumber
    \label{eq:FWH_unsteadyM}
\end{align}
Integration of Eq. \ref{eq:FWH} results in an unsteady mass addition (monopole) term corresponding to $Q$, an unsteady force (dipole) term corresponding to $F_i$, and an unsteady stress (quadrupole) term corresponding to $T_{ij}$. In our case, the flow speed is small and the volume integral term corresponding to $T_{ij}$ can be ignored. Only surface integrals are therefore required to be computed on a surface that encloses all acoustic sources in the problem. Such a surface is often called a Kirchoff surface. Figure \ref{fig:integration_surface} shows the integration surface used in the current predictions. It consists of a ``porous'' surface, labeled `Kirchoff surface 1' in the figure, which allows flow to pass through it, and an impermeable surface, labeled `Kirchoff surface 2'. 

Atmospheric absorption effects on sound propagation have to be accounted for in the numerical predictions to compare with experimental measurements. The atmospheric absorption coefficients are calculated using the ANSI standard \citep{ISO9613,bass1995atmospheric} for peak frequencies in the predicted spectrum with the temperature, relative humidity and pressure set as $293$ K, $50$\% and $1$ atm respectively.

\section*{Results and analysis of the baseline whistle}
\label{sec:results_active}
The baseline whistle is a Class III whistle with the resonating chambers acting as amplifiers. We aim to answer the following questions about this design: 1) what is the resonance mechanism at play (Helmholtz, Rossiter modes, pipe resonance, etc.)? 2) how does the fundamental frequency vary with supply air pressure/flow speed and resonator geometry, and 3) how does the sound intensity vary with these variables. A combined experimental-numerical approach is taken to perform a comprehensive analysis of the whistle. The experiments are conducted in the anechoic chamber at ISU and the numerical simulations are performed using the uRANS model described earlier.

\subsection*{Experimental results}
\label{sec:Exp_Baseline_Results}
\begin{figure}[htb!]
    \subcaptionbox{PSD spectrum \label{fig:Exp_Baseline_SmplSpec_PSD}}{\incfig[width=0.49\textwidth]{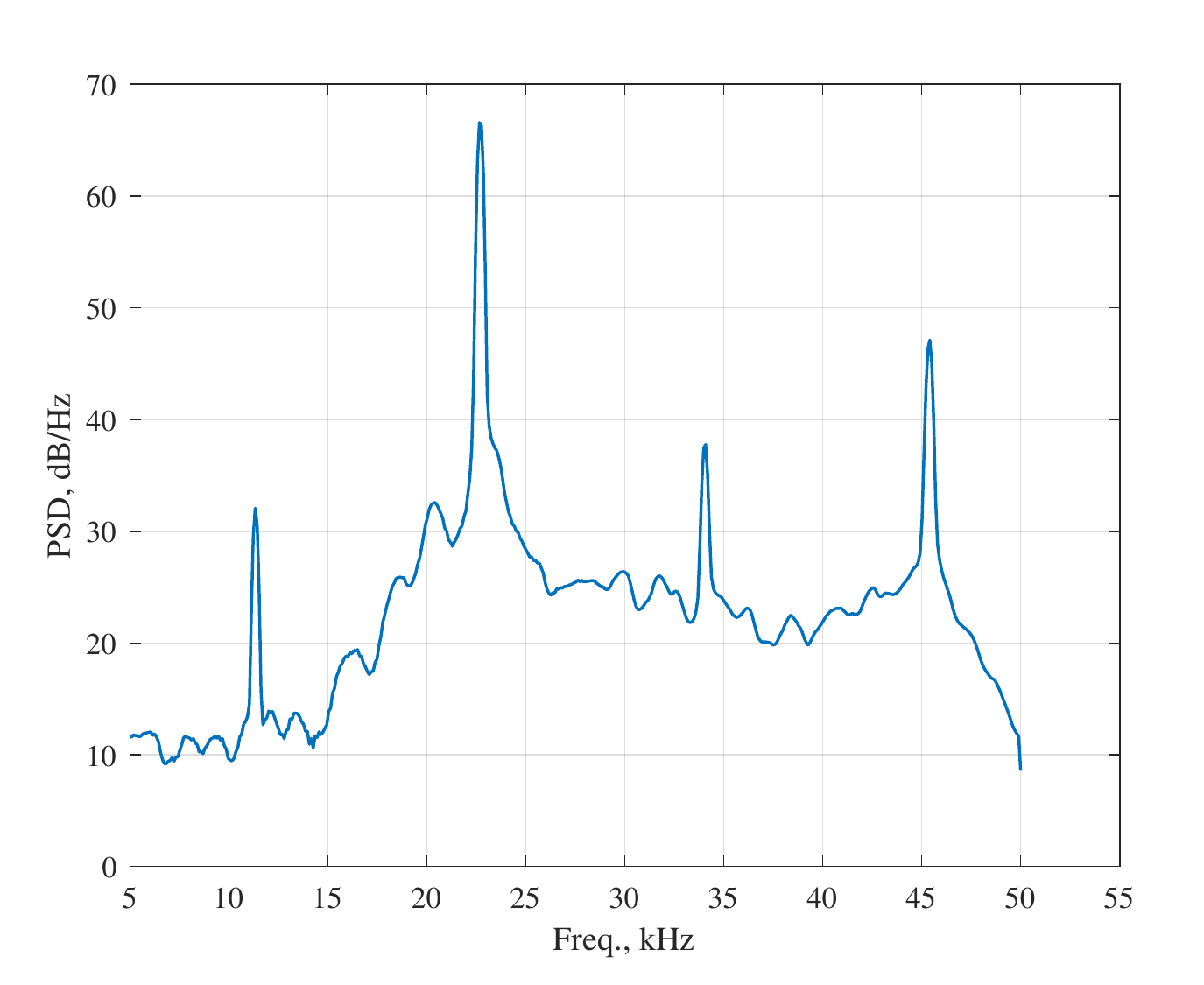}}
    \hfill
    \subcaptionbox{SPL of peak frequencies \label{fig:Exp_Baseline_SmplSpec_SPL}}{\incfig[width=0.49\textwidth]{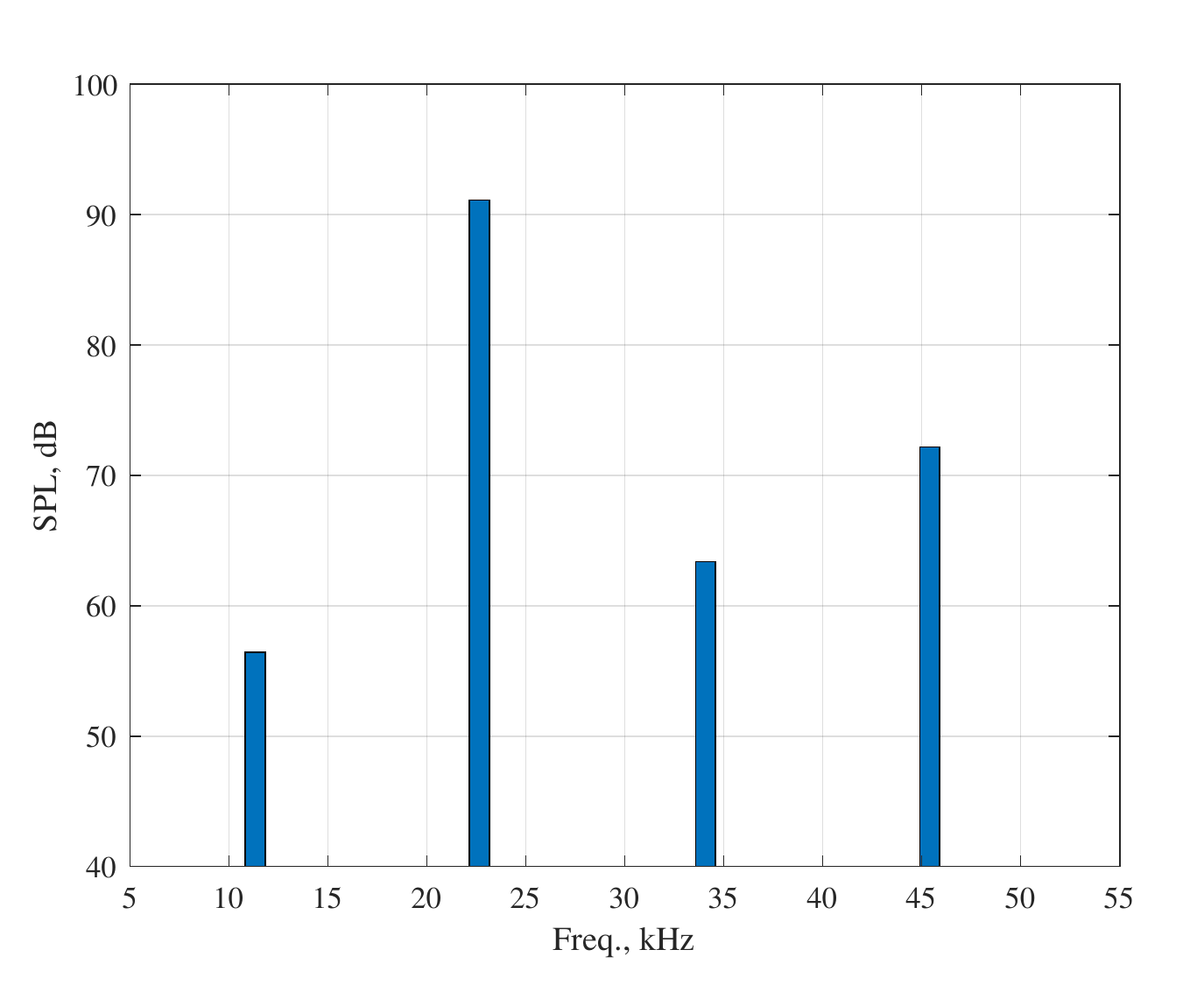}}
    \caption{Typical farfield sound spectrum of the baseline whistle, measured at supply pressure, $p_t = 2.0$ psig. (a) power spectral density (PSD) of pressure, and (b) sound pressure level (SPL) with reference to $20~\mu$Pa. The SPL computation requires integrating over a frequency band; the bandwidth used for integration is $2$ kHz.\label{fig:Exp_Baseline_SmplSpec}}
\end{figure}

Figure \ref{fig:Exp_Baseline_SmplSpec} shows the frequency spectrum of the farfield ultrasound radiation from the baseline whistle for a supply air pressure, $p_t=2.0$ psig. Figure \ref{fig:Exp_Baseline_SmplSpec_PSD} plots the power spectral density (PSD) of the pressure signal measured in the acoustic farfield of the whistle. The baseline whistle has a fundamental frequency around $23$ kHz; harmonics of the fundamental frequency are also observed in the spectrum. A sub-harmonic at around $\sim 11.5$ kHz is also measured. The PSD spectrum can be integrated over a finite frequency bandwidth ($=2$ kHz here) to obtain the tone sound pressure level (SPL). Figure \ref{fig:Exp_Baseline_SmplSpec_SPL} plots the SPL of the four tones measured in the frequency range $0 - 50$ kHz. The dominant peak at around $23$ kHz has an SPL of over $90$ dB (reference pressure is $20\mu$ pa).

\begin{figure}[htb!]
    \incfig[width=0.6\textwidth]{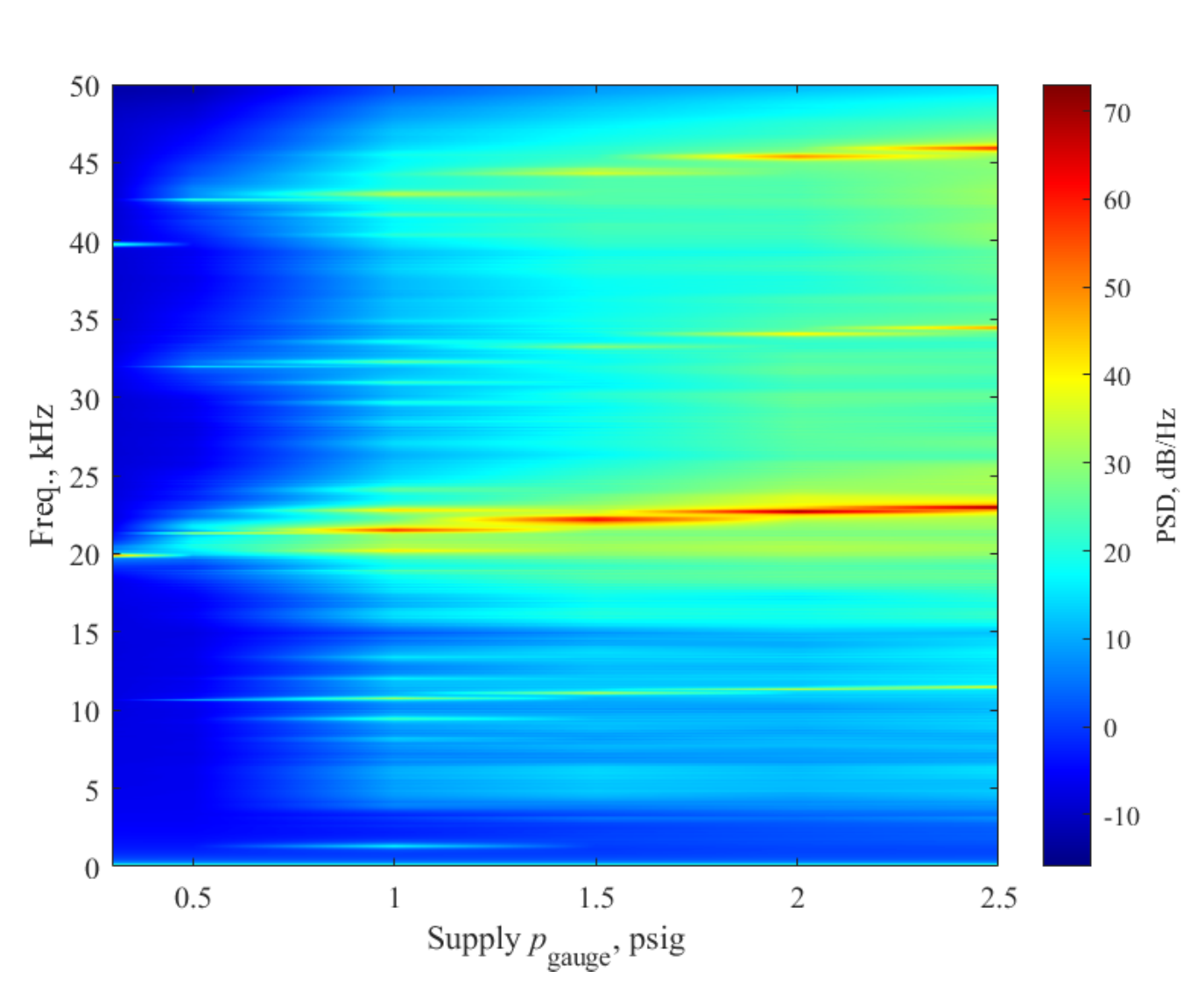}
    \caption{PSD spectra variation with supply air pressure ($p_t$); this can be viewed as a spectrogram showing variation with $p_t$ instead of time.}
    \label{fig:Exp_Baseline_PSDContour}
\end{figure}

The measured farfield sound spectra for $0.3 < p_t < 2.5$ psig are plotted in Fig. \ref{fig:Exp_Baseline_PSDContour}. Such plots are called spectrograms wherein variation of the frequency spectrum with time is considered; here, the variation is with $p_t$. In the pressure range considered here, the spectra show a fundamental (also peak) frequency between $20-24$ kHz. The peak frequency increases slightly with $p_t$. The first harmonic (between $40-48$ kHz) and a sub-harmonic ($10-12$  kHz) are also observed. 

\begin{figure}[htb!]
    \incfig[width=0.7\textwidth]{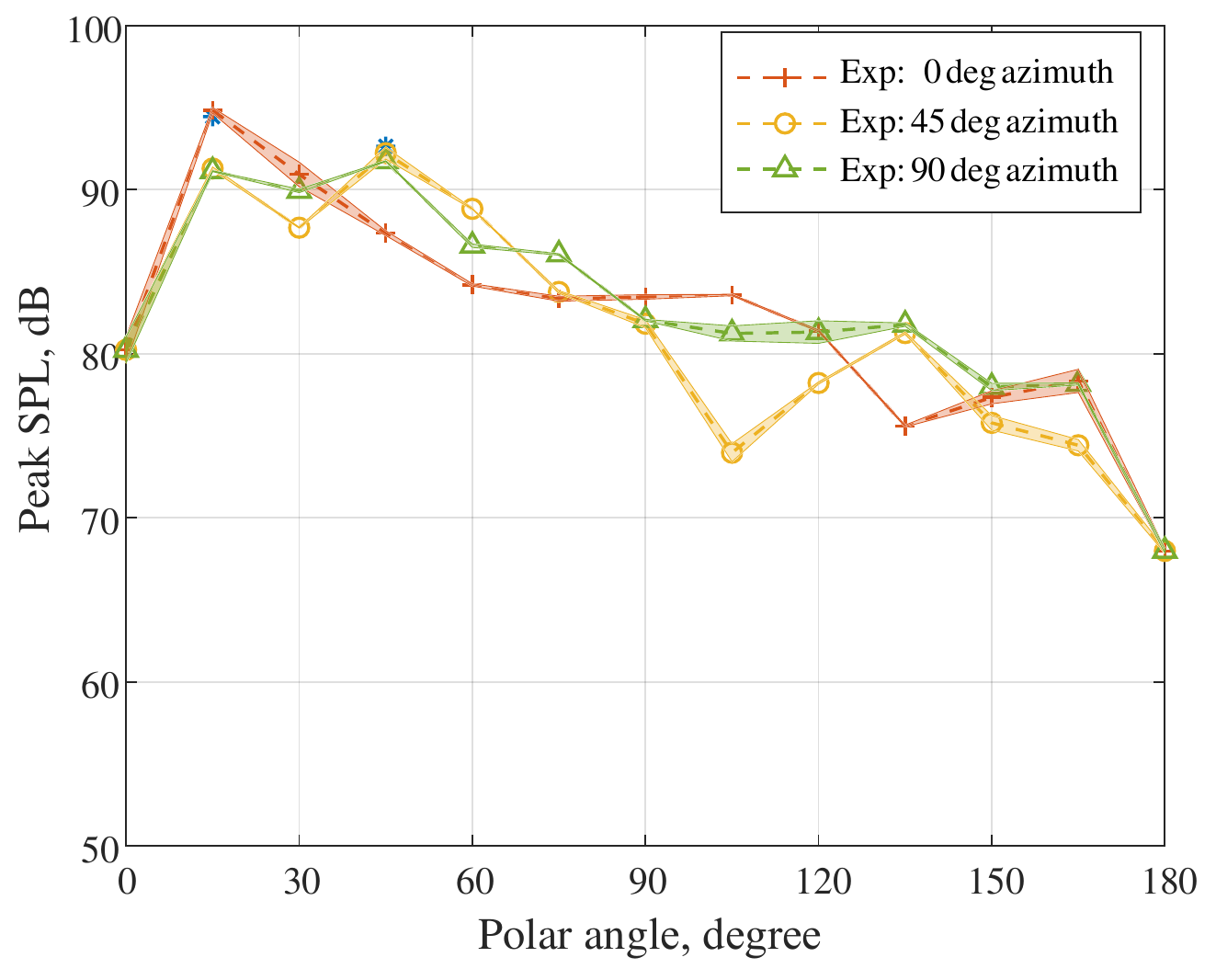}
    \caption{Directivity of sound radiation from the baseline whistle for three azimuthal angles - $0^\circ,~45^\circ,~\&~90^\circ$. Also shown with thin lines around each reading are the error bounds (one standard deviation) computed from five separate measurements.
    \label{fig:exp_directivity}}
\end{figure}

High-frequency sound can be highly directional. An investigation of the directionality of sound radiation from the baseline whistle is conducted for one value of supply pressure, $p_t=2.0$ psig. Polar directivity is examined by measuring the sound signal along a $1.651$ m arc and azimuthal directivity is assessed by measuring at $0^\circ$, $45^\circ$ and $90^\circ$ azimuthal angles. Figure \ref{fig:exp_directivity} plots the directivity of the SPL of the peak frequency. As expected, the radiated sound is highly directional with maximum radiation occurring between polar angles $15^\circ - 45^\circ$; note that the polar angle is measured from downstream of the whistle with the whistle exhaust pointing at $0^\circ$. The azimuthal variation (compare the three dashed lines with symbols in the figure) is much smaller although not negligible.
Five readings were taken to quantify the measurement uncertainty. The upper and lower bounds of the five readings are also plotted in Fig. \ref{fig:exp_directivity} with thin lines of the same color. The measurement uncertainty ($\pm 0.5~\times$ standard deviation) computed using five separate readings is shown with the shaded region bounded by thin solid lines.
 
\subsection*{Two-dimensional CFD analysis}
\label{sec:2D_CFD}
\begin{figure}[htb!]
    \subcaptionbox{2D-CFD domain \label{fig:2D-CFD_Mesh_general}}{\incfig[width=0.24\textwidth]{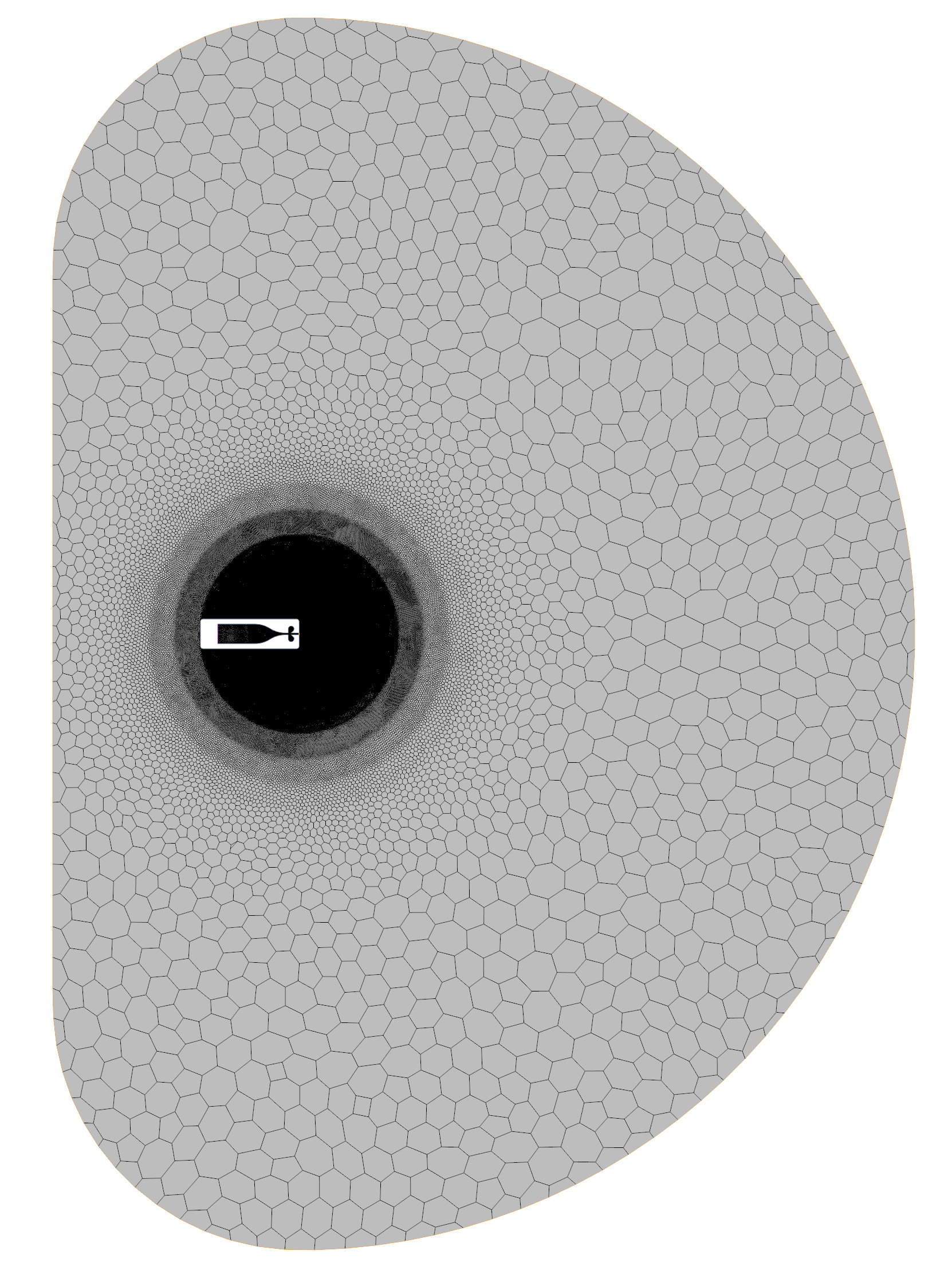}}
    \hfill
    \subcaptionbox{whistle-interior mesh
    \label{fig:2D-CFD_Mesh_whistle}}{\incfig[width=0.49\textwidth]{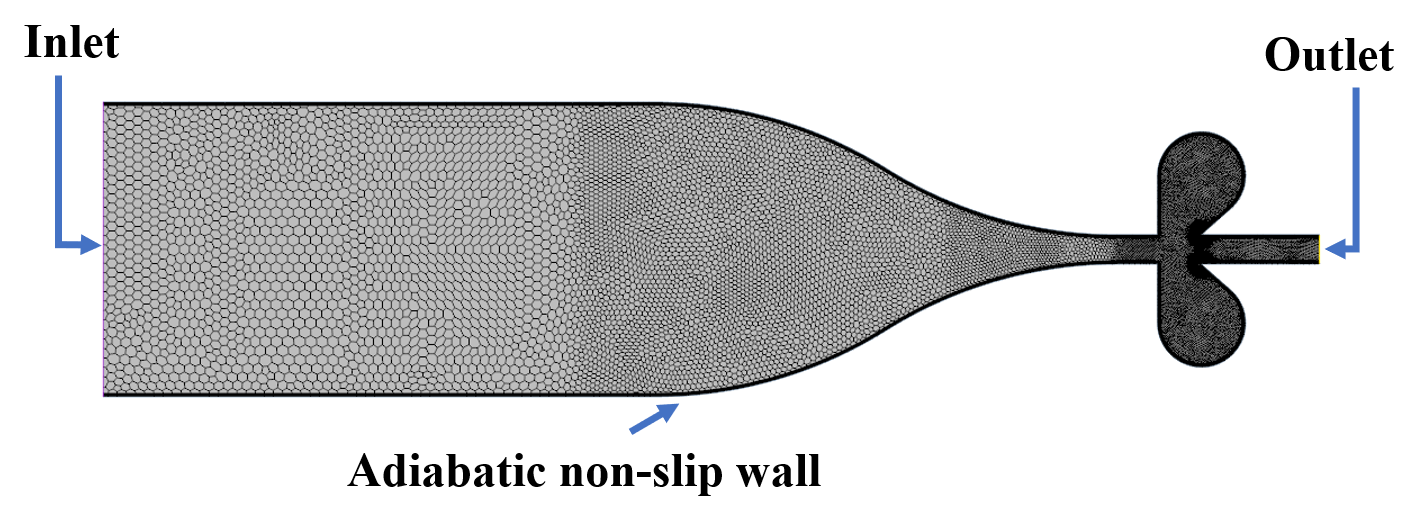}}
    \hfill
    \subcaptionbox{interface
    \label{fig:2D-CFD_Mesh_interface}}{\incfig[width=0.24\textwidth]{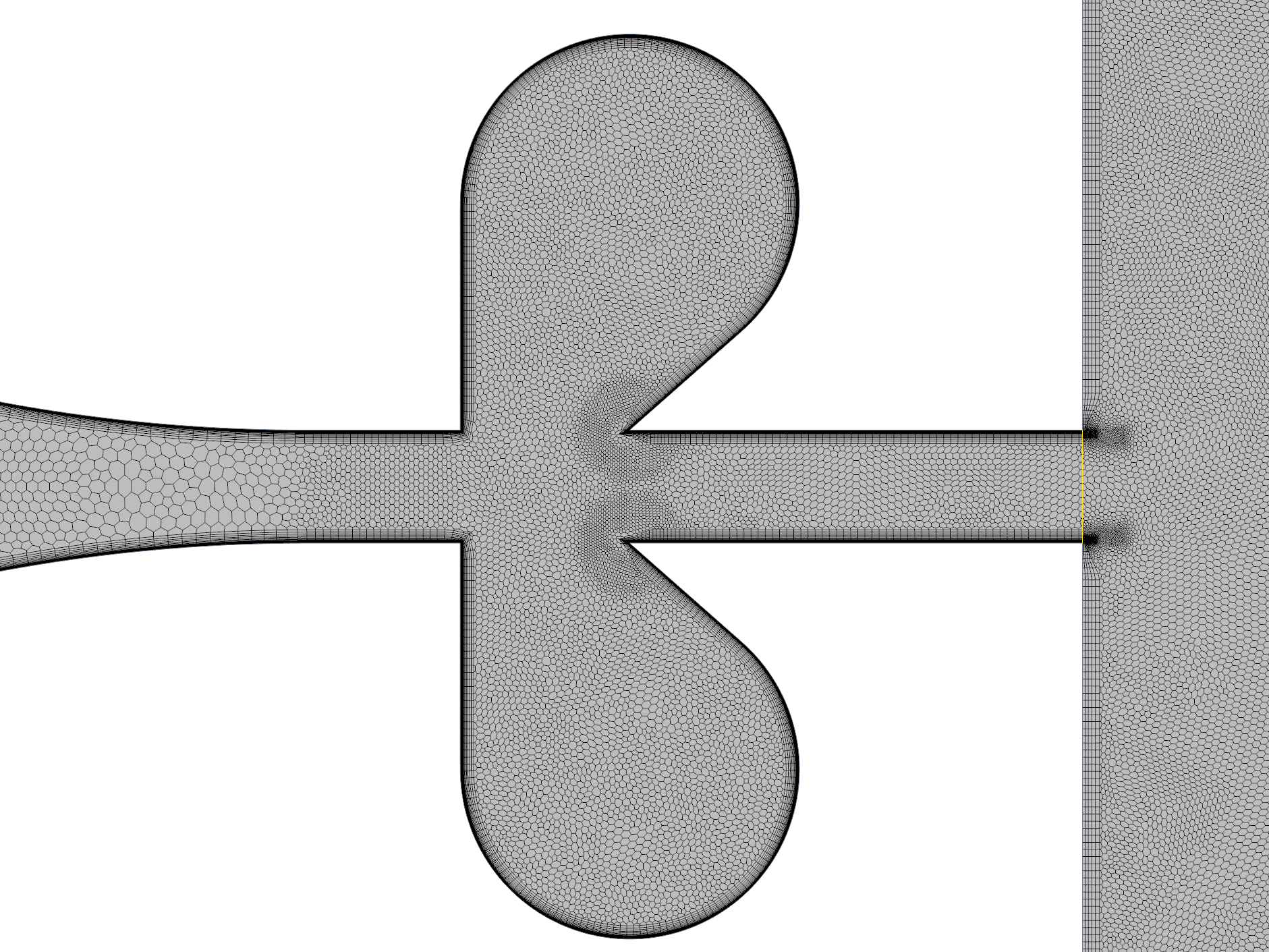}}
    \caption{The computational mesh used for 2D simulations. (a) overall domain showing the large absorbing chamber, (b) mesh inside the whistle, and (c) zoom view of the mesh in the resonators and near the whistle exit.
    \label{fig:2D-CFD_Mesh}}
\end{figure}

To supplement the farfield acoustic measurements with flow field information, two-dimensional, unsteady fluid flow simulations are performed. The uRANS equations are solved with the $k-\omega$ SST turbulence model in STAR-CCM+. Figure \ref{fig:2D-CFD_Mesh} shows the CFD domain, the mesh and the boundary conditions used in the simulations. Adiabatic and no-slip wall conditions are applied on the side boundaries. Stagnation pressure is prescribed at the inlet boundary and the outlet boundary is handled as follows. A large region (absorbing chamber) at the nozzle exit is meshed to allow for acoustic wave radiation out of the whistle (see Fig. \ref{fig:2D-CFD_Mesh_general}); the boundaries of the large region are prescribed to be at ambient pressure. Figure \ref{fig:2D-CFD_Mesh_interface} provides a detailed view of the resonator and the whistle interface with the absorbing chamber.

\begin{figure}[!htb]
    \centering
    \subcaptionbox{probe locations\label{fig:2D-CFD_probes}}{\incfig[width=0.45\textwidth]{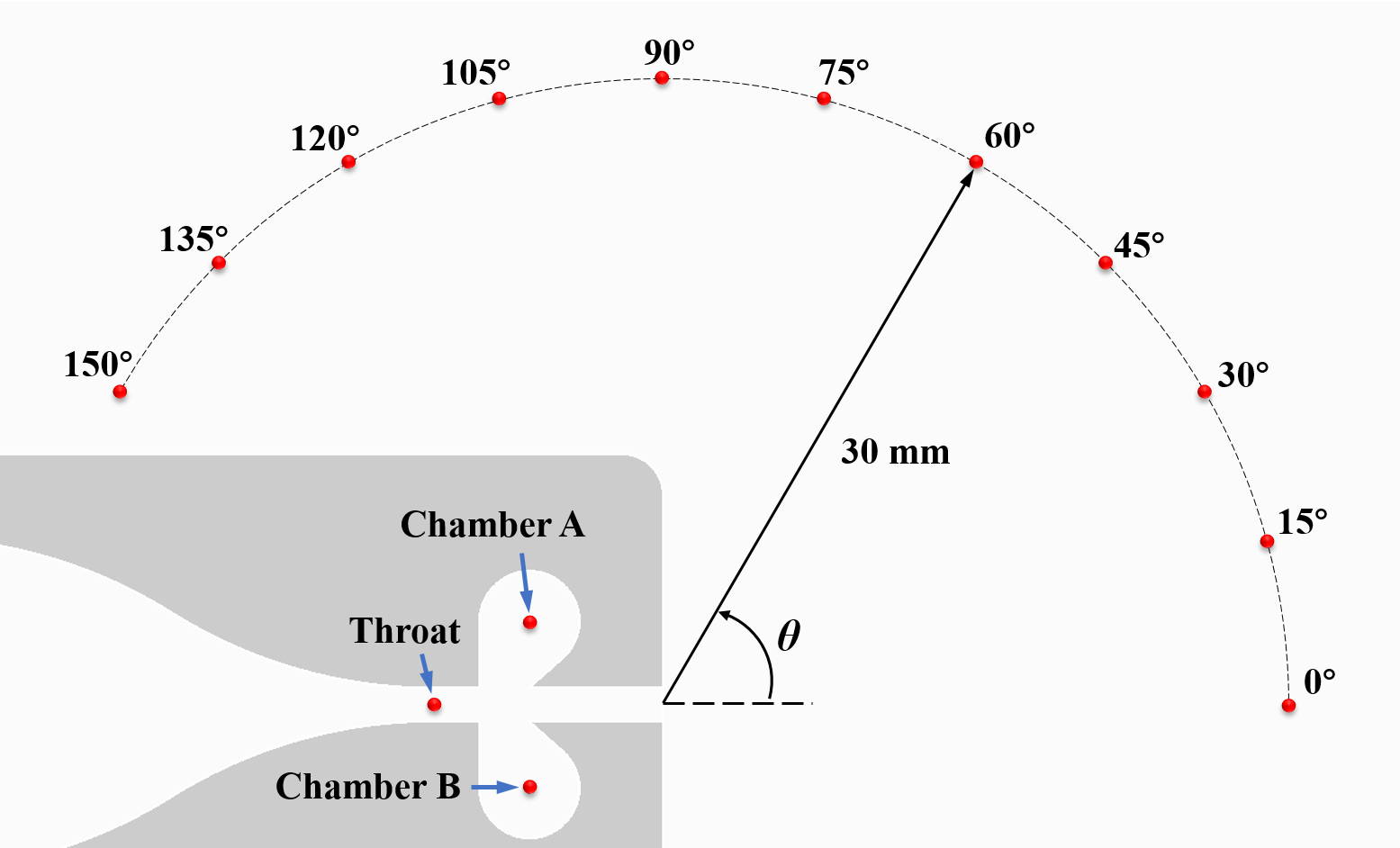}}
    \hfill
    \subcaptionbox{instantaneous pressure contours\label{fig:2DCFD_Baseline_PfieldRadiate}}{\incfig[width=0.50\textwidth]{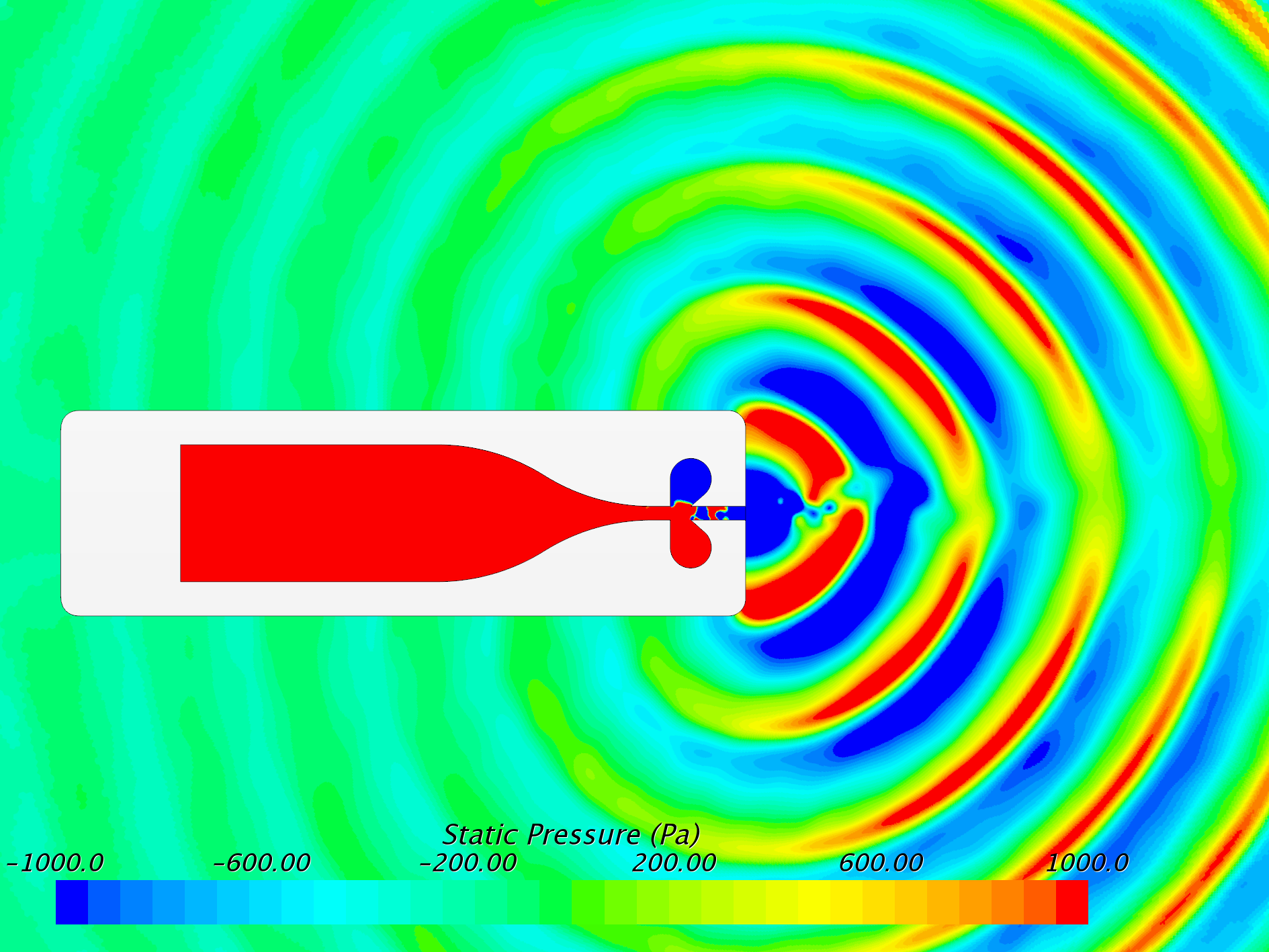}}
    \caption{Two-dimensional CFD analysis. (a) Probe locations where time-accurate data is collected for spectral analysis, and (b) visualization of the acoustic wave fronts via pressure contours.}
    \label{fig:2D-CFD}
\end{figure}

Time accurate data is collected at several (probe) locations inside the whistle and in the absorbing chamber. This data is processed to obtain the ultrasound spectra and directivity. Inside the whistle, a probe is placed along the centerline of the flow channel at the throat (inlet of the resonator section), and two probes are placed at the center of each resonating chamber. Outside the whistle, seven probes are placed in a circular arc of radius $30$ mm centered at the whistle exit. The probe locations are illustrated in Fig. \ref{fig:2D-CFD_probes}.

\begin{figure}[!htb]
    \subcaptionbox{PSD of radiated sound at 30$^\circ$ polar angle\label{fig:2dCFD_Spec_30mm30deg}}{\incfig[width=0.45\textwidth]{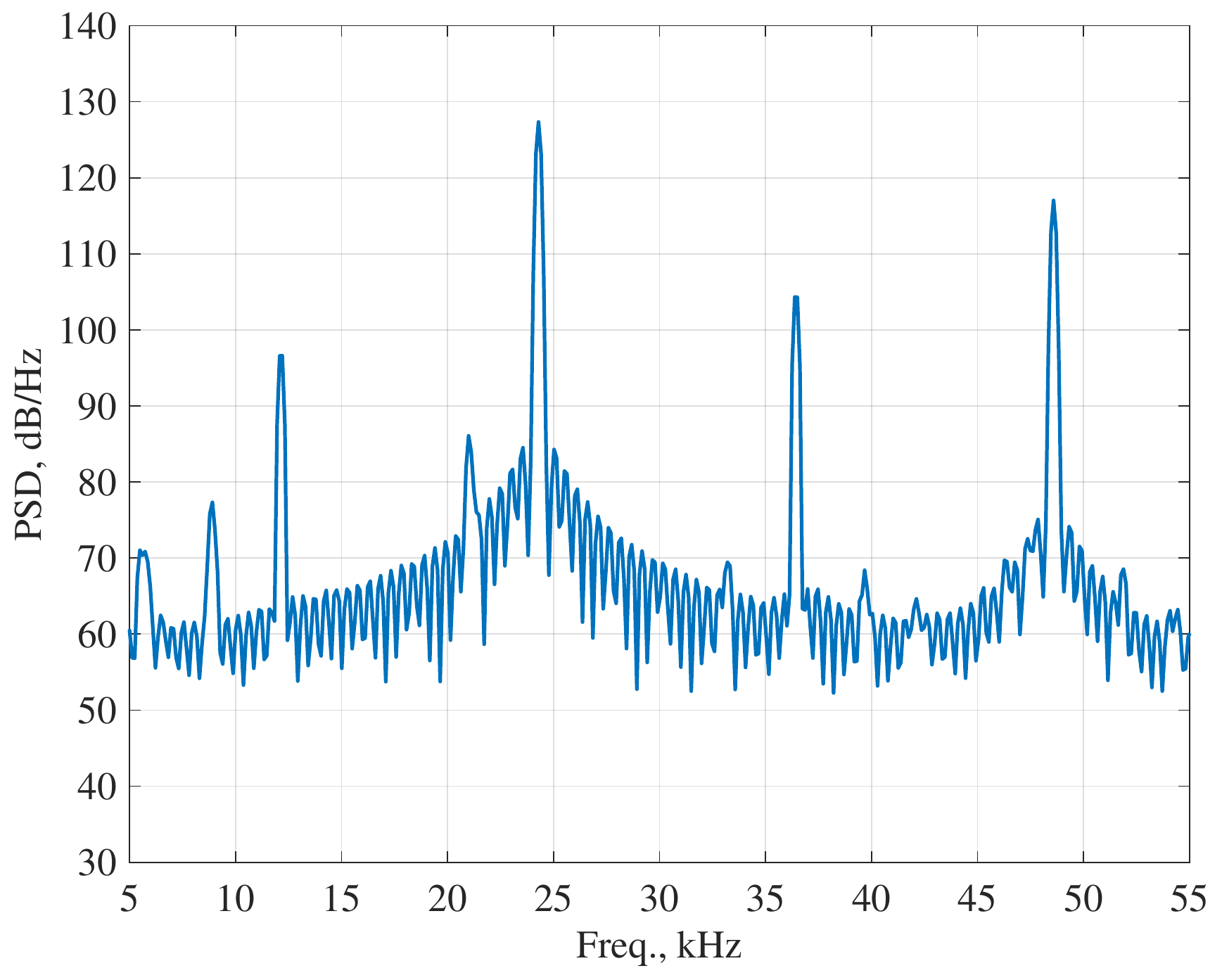}}
    \hfill
    \subcaptionbox{directivity of radiated sound\label{fig:2dCFD_polar}}{\incfig[width=0.55\textwidth]{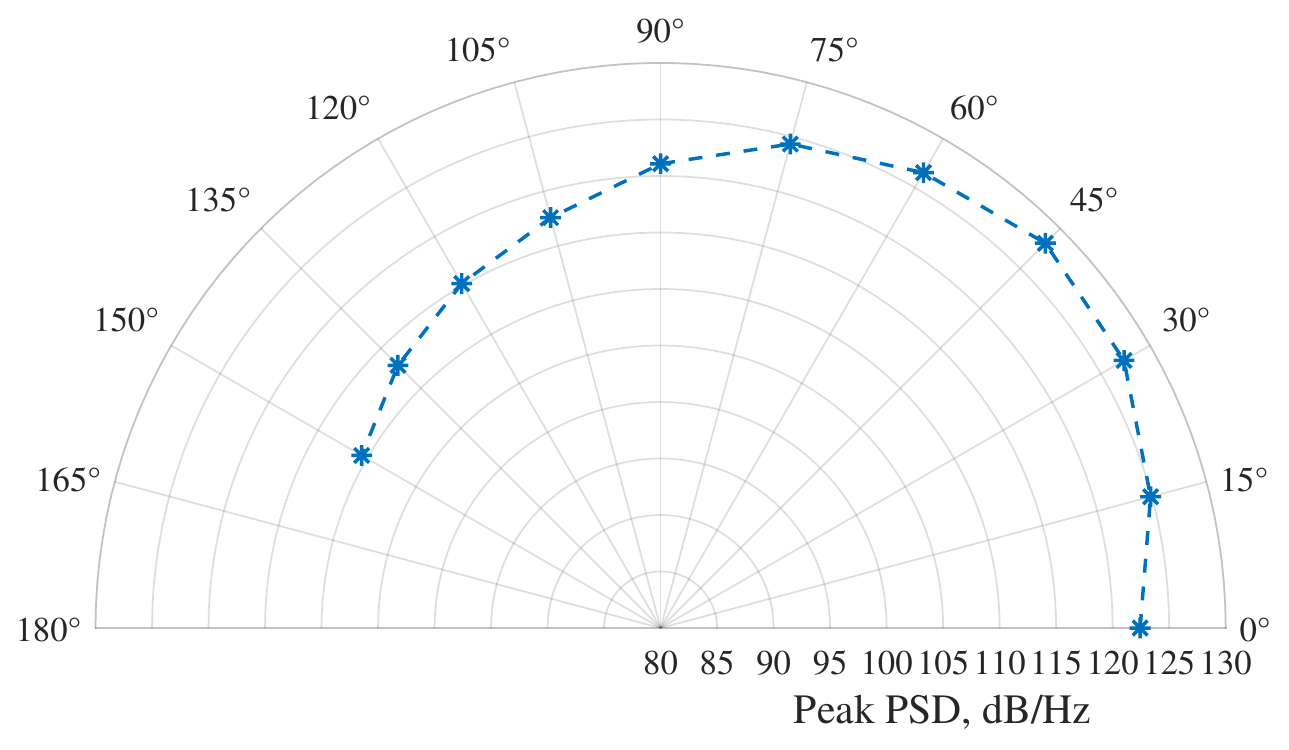}}
    \caption{2D CFD results of the baseline whistle for $p_t=2.0$ psig. (a) PSD of radiated pressure field, and (b) farfield sound directivity
    \label{fig:2dCFD_spec_directivity}}
\end{figure}

Simulations are conducted over a range of inlet stagnation pressures ($0.1\leq p_{t}\leq 2.5~\rm psig$). The simulation result for $p_t = 2.0$ psig is discussed in detail here. Figure \ref{fig:2DCFD_Baseline_PfieldRadiate} shows the radiating sound field. Figure \ref{fig:2dCFD_Spec_30mm30deg} plots the power spectral density of the acoustic pressure in the farfield. The peak frequency of the radiating sound field is $\sim 24$ kHz. Harmonics and sub-harmonics of the peak frequency are also observed. The predicted spectrum qualitatively agrees with the experimental result in Fig. \ref{fig:Exp_Baseline_SmplSpec_PSD}.
 
The directionality of the radiated sound field can be inferred from Fig. \ref{fig:2DCFD_Baseline_PfieldRadiate}. Figure \ref{fig:2dCFD_polar} is a polar plot of the pressure PSD of the peak frequency, which shows the peak radiation to be occurring at around $45^\circ$ polar angle. This qualitatively agrees with the measurements (compare with Fig. \ref{fig:exp_directivity}, which shows peak radiation off-axis, between $15^\circ - 45^\circ$ polar angle). 

\begin{figure}[!htb]
    \subcaptionbox{at a chamber center\label{fig:1dCFD_Spec_chamberA}}{\incfig[width=0.49\textwidth]{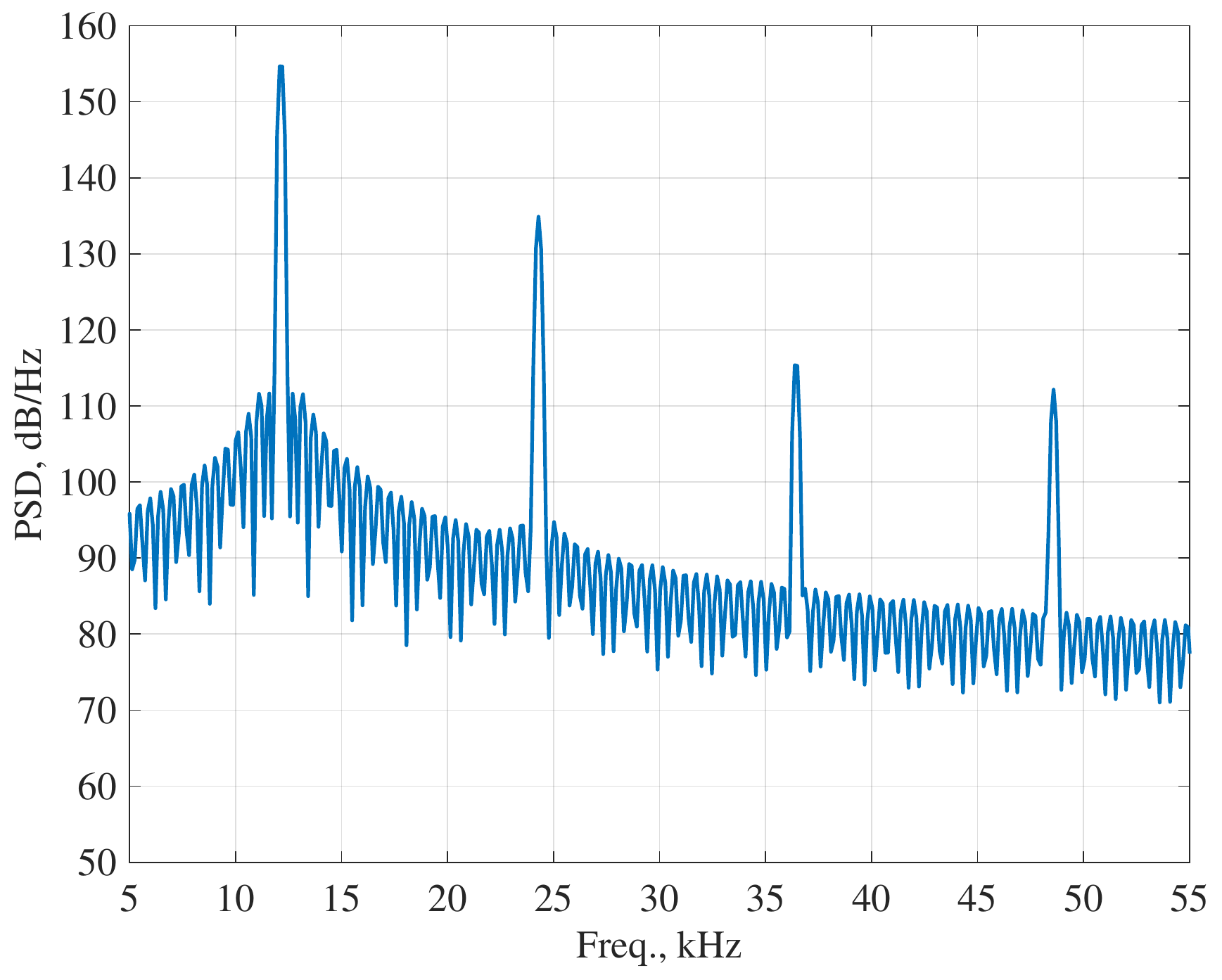}}
    \hfill
    \subcaptionbox{in the farfield}{\incfig[width=0.49\textwidth]{Figs/2DCFD_Baseline_2psig_30mm30deg.pdf}}
    \caption{Comparison of pressure PSDs measured (a) at the center of one of the resonating chambers 
    and (b) in the farfield
    \label{fig:2dCFD_SpecSample}}
\end{figure}

\begin{figure}[htb!]
    \subcaptionbox{Pressure field in the resonators \label{fig:2DCFD_ChamberP_Pfield}}{\incfig[width=0.48\textwidth]{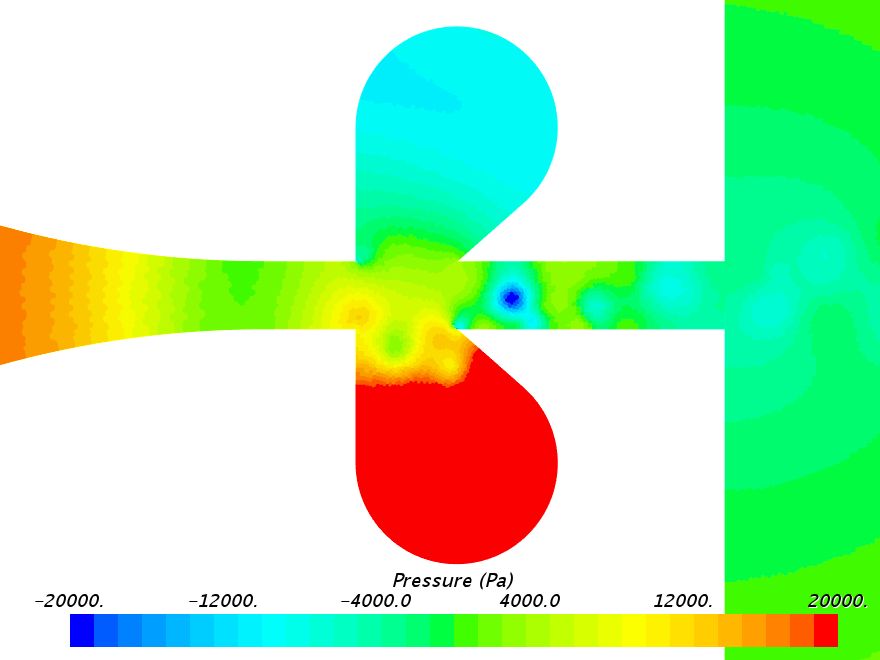}}
    \hfill
    \subcaptionbox{temporal variation of pressure in the chambers
    \label{fig:2DCFD_ChamberP_tSig}}{\incfig[width=0.48\textwidth]{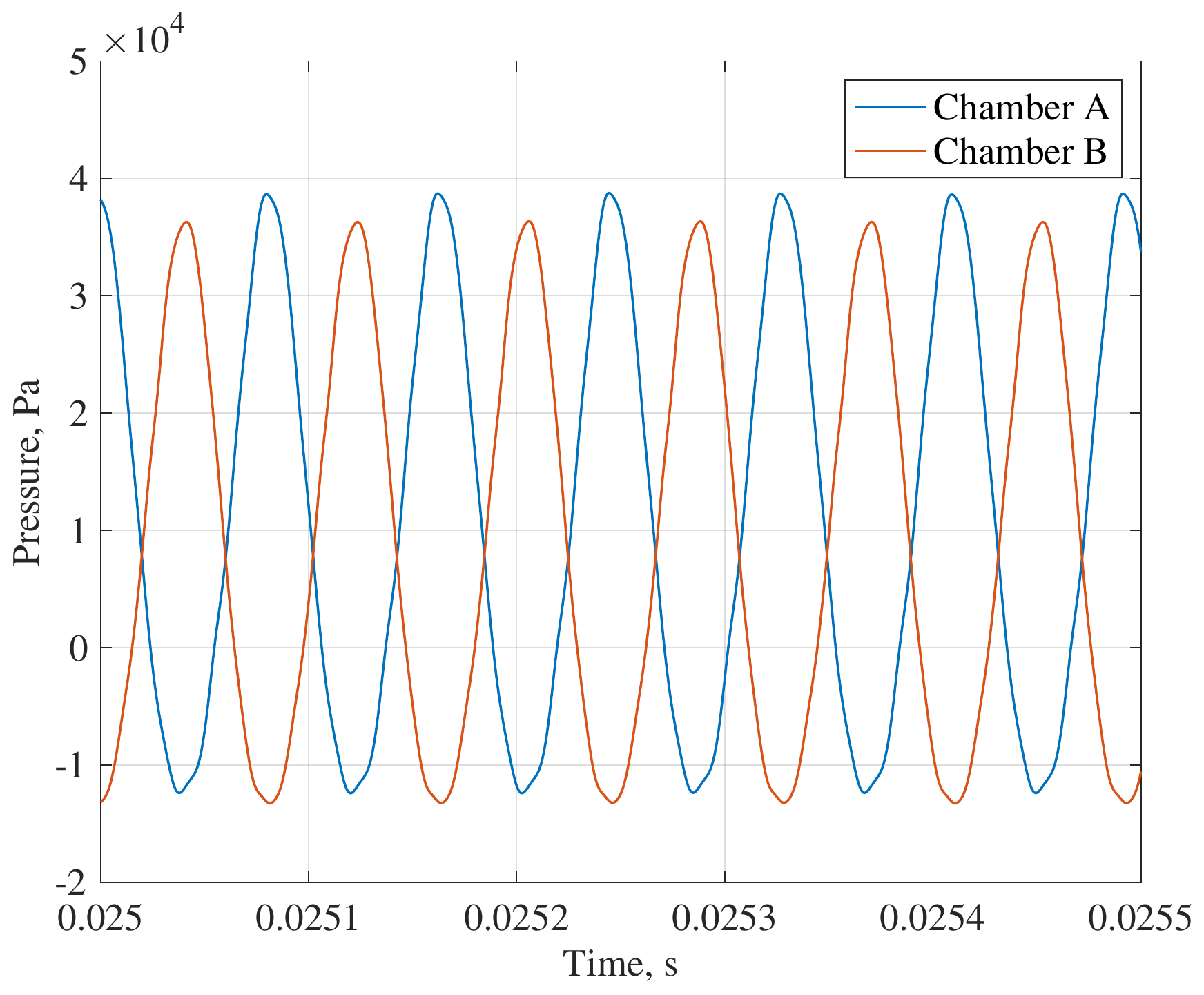}}
    \caption{Out-of-phase pressure oscillation in the two chambers/resonators of the whistle
    \label{fig:2DCFD_ChamberP}}
\end{figure}

Figure \ref{fig:2dCFD_SpecSample} contrasts the pressure PSD spectra at the center of Chamber A with that in the farfield. The acoustic signal at the chamber center has a peak frequency of $12$ kHz, which is half of the peak frequency of the radiating sound. Figure \ref{fig:2DCFD_ChamberP_Pfield} provides a snapshot of the pressure field inside the two resonators and Fig. \ref{fig:2DCFD_ChamberP_tSig} compares the time histories of the pressure at the center of the two chambers, which conclusively shows that the two chambers are operating out-of-phase. These results indicate that the basic mechanism of the baseline whistle is as follows. The chambers/resonators have a fundamental frequency around $12$ kHz. Since the chambers operate out-of-phase, the fundamental (and the odd harmonics) of the two chambers cancel each other, and the even harmonics radiate to the far field. The second harmonic ($\sim 24$ kHz) therefore appears in the farfield as the peak frequency. Since the phase cancellation is not perfect, some energy radiates in the fundamental and the odd harmonics.

We next investigate the performance of the baseline whistle with varying inlet (supply) pressure and geometry. 
%
\paragraph{Variation with supply pressure}
\label{subsec:2DCFD_multiPt}
%
Two-dimensional CFD simulations of the baseline whistle are performed for several $p_t$ values in the range $0.1 \leq p_t \leq 2.5~\rm psig$. The results are summarized in Table \ref{tab:VariousPt}. 
\begin{table}[htbp!]
  \centering
  \caption{Summary of 2D CFD simulation results for varying $p_t$\label{tab:VariousPt}}
    \begin{tabular}{c|c|c|c|c}
    $p_t$ & \textit{avg.} $M$ at \textit{throat} & \textit{peak chamber freq.} & \textit{peak farfield freq.} & \textit{phase relation} \\
    (psig) & & (kHz) & (kHz) & \textit{between chambers} \\
     \hline \hline
    0.1 & 0.085 & 6.10 & 12.21 & in-phase \\
    0.3 & 0.142 & 9.77 & 19.65 & out-of-phase \\
    0.5 & 0.172 & 11.11 & 22.09 & out-of-phase \\
    0.8 & 0.165 & 11.11 & 21.36 & out-of-phase \\
    1.0 & 0.178 & 10.99 & 21.85 & out-of-phase \\
    1.2 & 0.162 & 11.11 & 22.09 & out-of-phase \\
    1.4 & 0.167 & 11.23 & 22.58 & out-of-phase \\
    1.6 & 0.169 & 11.47 & 22.83 & out-of-phase \\
    1.8 & 0.175 & 11.60 & 23.32 & out-of-phase \\
    2.0 & 0.163 & 12.08 & 24.29 & out-of-phase \\
    2.1 & 0.164 & 12.21 & 24.29 & out-of-phase \\
    2.3 & 0.163 & 12.21 & 24.41 & out-of-phase \\
    2.5 & 0.163 & 12.21 & 24.54 & out-of-phase \\
    \hline \hline
    \end{tabular}%
\end{table}%
 
\begin{figure}[htb!]
    \subcaptionbox{chamber center\label{fig:2DCFD_PSDContour_30mm30deg _chamberA}}{\incfig[width=0.5\textwidth]{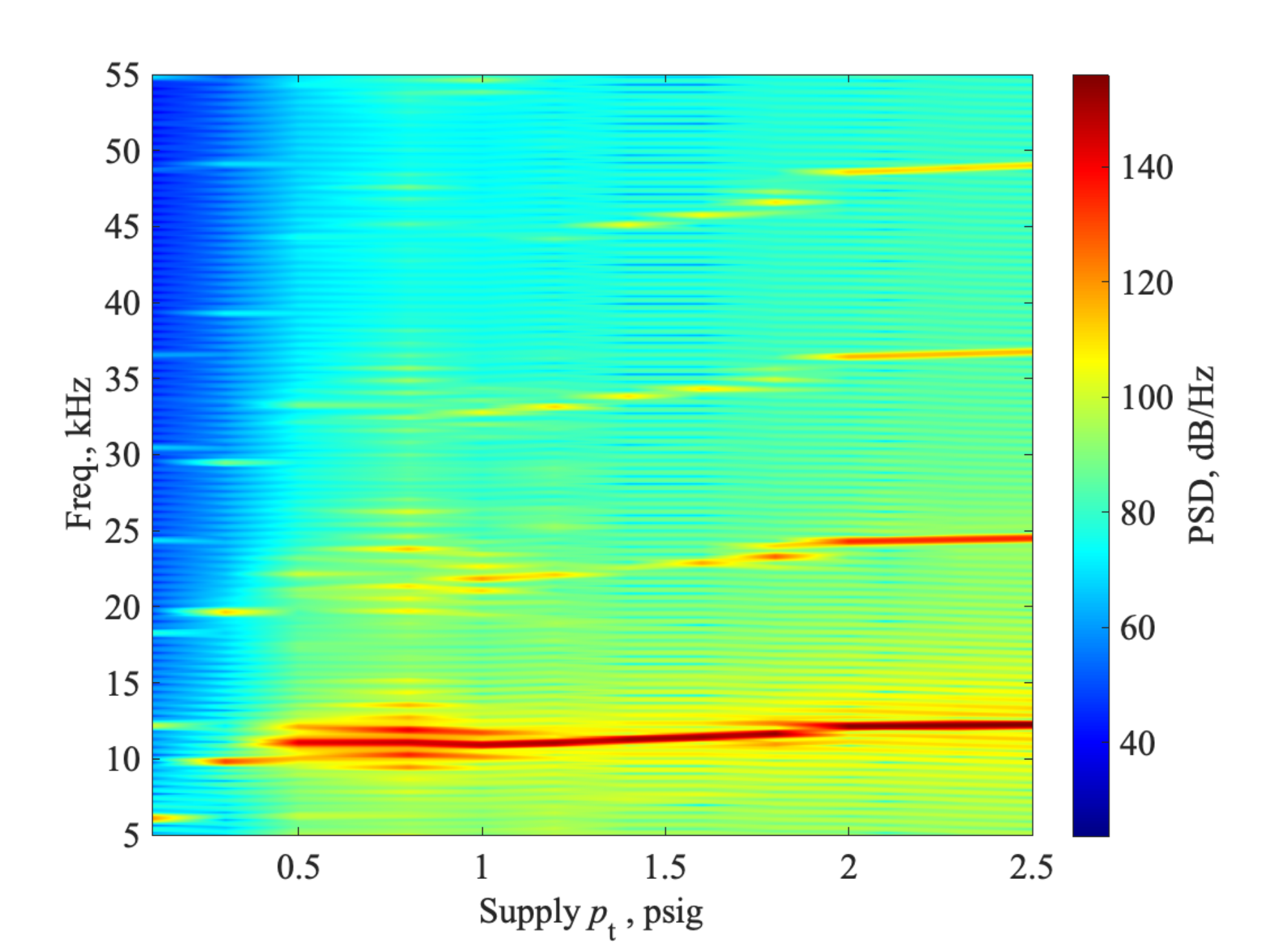}}
    \hfill
    \subcaptionbox{radiated ultrasound in the farfield\label{fig:2DCFD_PSDContour_30mm30deg}}{\incfig[width=0.47\textwidth]{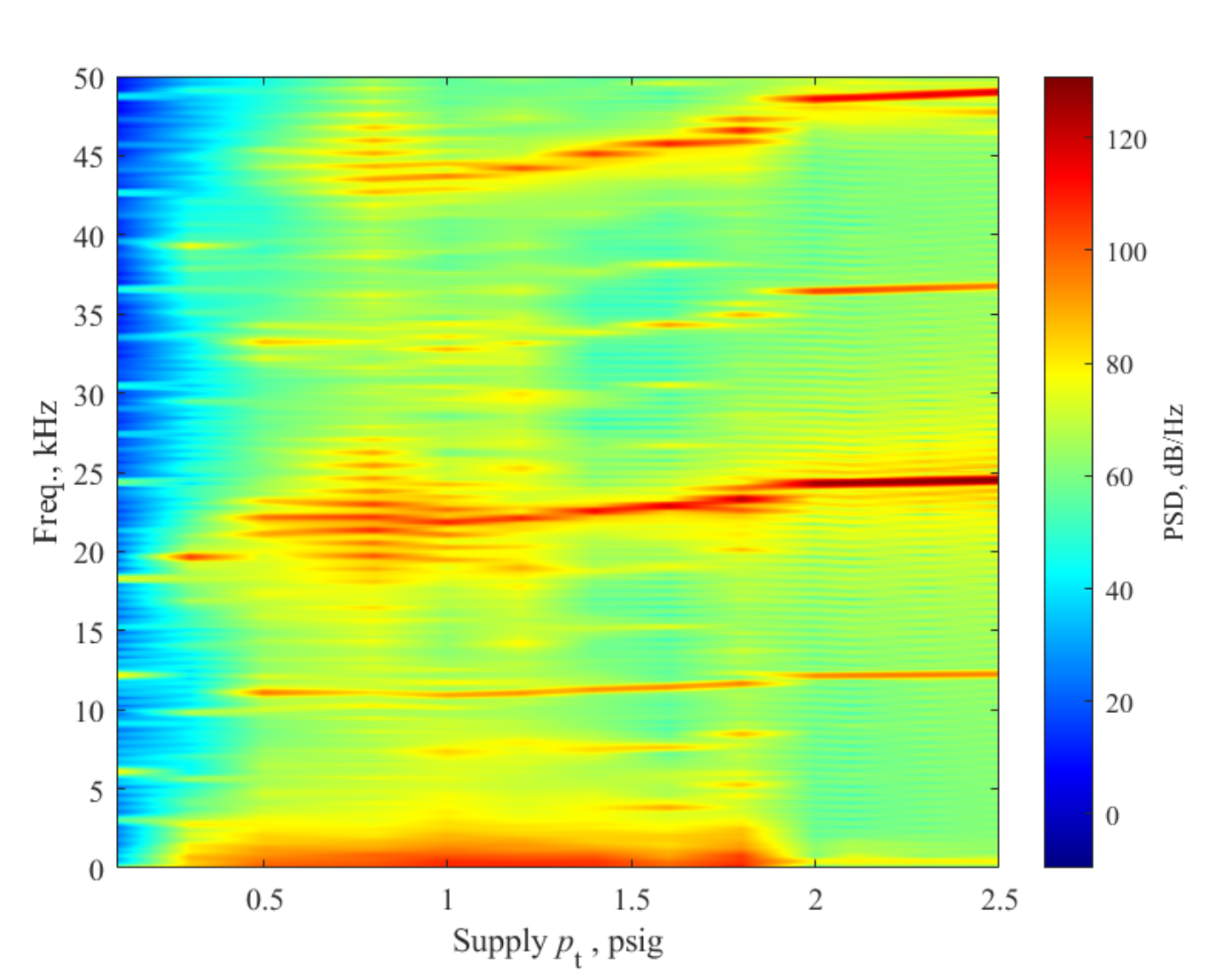}}
    \caption{2D CFD results of the baseline whistle. Variation of PSD spectra with $p_t$: (a) at the center of one of the chambers, and (b) $30$ mm away from the whistle (in the farfield) at $30^\circ$ polar angle.\label{fig:2dCFD_PSDContour}}
\end{figure}

Variations with $p_t$ of pressure PSD spectra in the farfield and at the center of one of the resonating chambers are plotted as ``spectrograms'' in Fig. \ref{fig:2dCFD_PSDContour}. The numerical results in Fig. \ref{fig:2DCFD_PSDContour_30mm30deg} are qualitatively similar to the corresponding experimental measurements in Fig. \ref{fig:Exp_Baseline_PSDContour}. Both show the peak frequency to be around $23$ kHz, which increases slightly with $p_t$. In the results from both methods, there appears to be a threshold $p_t$ value below which the whistle does not produce any tones, or is ``cut-off''.

%
\begin{figure}[htb!]
    \incfig[width=0.5\textwidth]{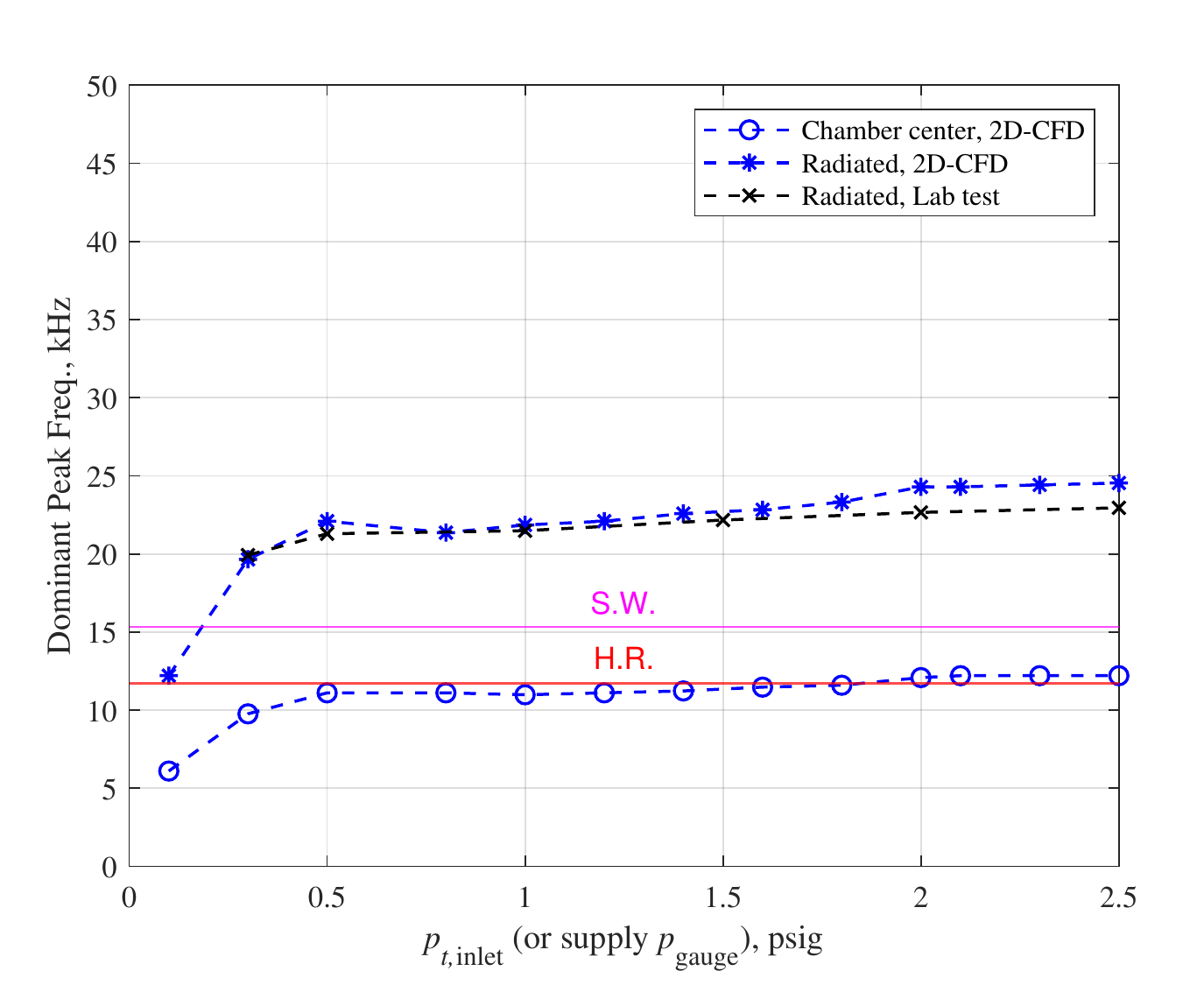}
    \caption{Variations with $p_t$ of peak frequencies in the farfield obtained from 2D CFD and measurements and in one of the chambers (from CFD). Also plotted are estimated Helmholtz resonance (H.R.) and standing-wave (S.W.) resonance frequencies.
    \label{fig:freq_vs_pt}}
\end{figure}

Figure \ref{fig:freq_vs_pt} compares the variations with $p_t$ of peak frequencies at the center of one of the resonating chambers with that of the radiating farfield sound. Also plotted in the figure are curves that correspond to the frequencies associated with different sound generation mechanisms that are theoretically possible in the baseline whistle. These include Helmholtz resonance (H.R.) and standing wave (S.W.) resonance in the chamber. The closed-form expressions relating these frequencies to geometric and flow parameters are provided in Appendix A.
 
Figure \ref{fig:freq_vs_pt} shows that the peak frequency in the resonating chambers closely matches the theoretical Helmholtz resonance frequency for this resonator geometry. Also, the frequency of the first Rossiter mode (cavity resonance) coincides with the H.R. for this design over the $p_t$ range considered here (see Appendix A). Once resonance is established, the resonance-amplified signal (feedback) controls the flow instability and dictates the shear layer flapping frequency \citep{gloerfelt2009cavity}. The observations therefore suggest that in the range of $p_t$ evaluated here, the device operates as a Class III whistle with Helmholtz resonance amplifying the acoustic signal. To differentiate between Helmholtz resonance and Rossiter modes, we investigate a different whistle design where these frequencies are distinct. Appendix B presents the results of this study where we found that the frequency observed in the chamber is close to the theoretical estimate of the Helmholtz resonance frequency and not that of Rossiter modes.

Since the two chambers (resonators) operate out of phase, the sound radiated by the whistle into the farfield is at double the Helmholtz resonance frequency, as seen in Fig. \ref{fig:freq_vs_pt}. The peak frequency increases slightly with $p_t$, which can be caused by the change in the geometry (likely the ``effective'' neck) of the resonator due to the variation in boundary layer thickness with $p_t$.  At small $p_t$ ($0.1 \le p_t \le 0.5$ psig), the peak frequency and the throat Mach number ($M$) increase with $p_t$. However, once $p_t$ exceeds $0.5$ psig, these quantities become insensitive to $p_t$. Note that while the peak frequency and $M$ do not vary with $p_t$ in this range, the peak SPL increases with increasing $p_t$.
%

\paragraph{Variation with whistle size}
\label{sec:2DCFD_multisize}
The peak frequency of the whistle can be altered by geometrically scaling the resonators. To target multiple frequencies in the $20$ kHz to $50$ kHz range, we investigate multiple designs which are obtained by geometrically scaling down the baseline whistle. The percentage scaling (based on linear dimension) considered here and the corresponding peak frequencies and SPL (obtained using 2D CFD) are tabulated in Table \ref{tab:VariousSize}.
\begin{table}[htbp]
  \centering
  \caption{Peak frequencies and SPLs of several whistles obtained by geometrically scaling down the baseline whistle. The peak frequency and SPL are obtained from pressure signal captured at $30^\circ$ polar angle, $30$ mm away from the whistle exit.
  \label{tab:VariousSize}}
    \begin{tabular}{c|c|c}
    \textit{size (\% of baseline)} & \textit{peak freq.} (kHz) & \textit{peak SPL} (dB) \\
    \hline\hline
    100& 24.29 & 151.34\\
    85 & 28.44 & 150.31\\
    70 & 34.55 & 149.01\\
    60 & 40.16 & 146.60\\
    55 & 43.82 & 145.22\\
    50 & 48.10 & 143.02\\
    \hline\hline
    \end{tabular}%
\end{table}%

\begin{figure}[!htb]
    \centering
    \incfig[width=0.6\textwidth]{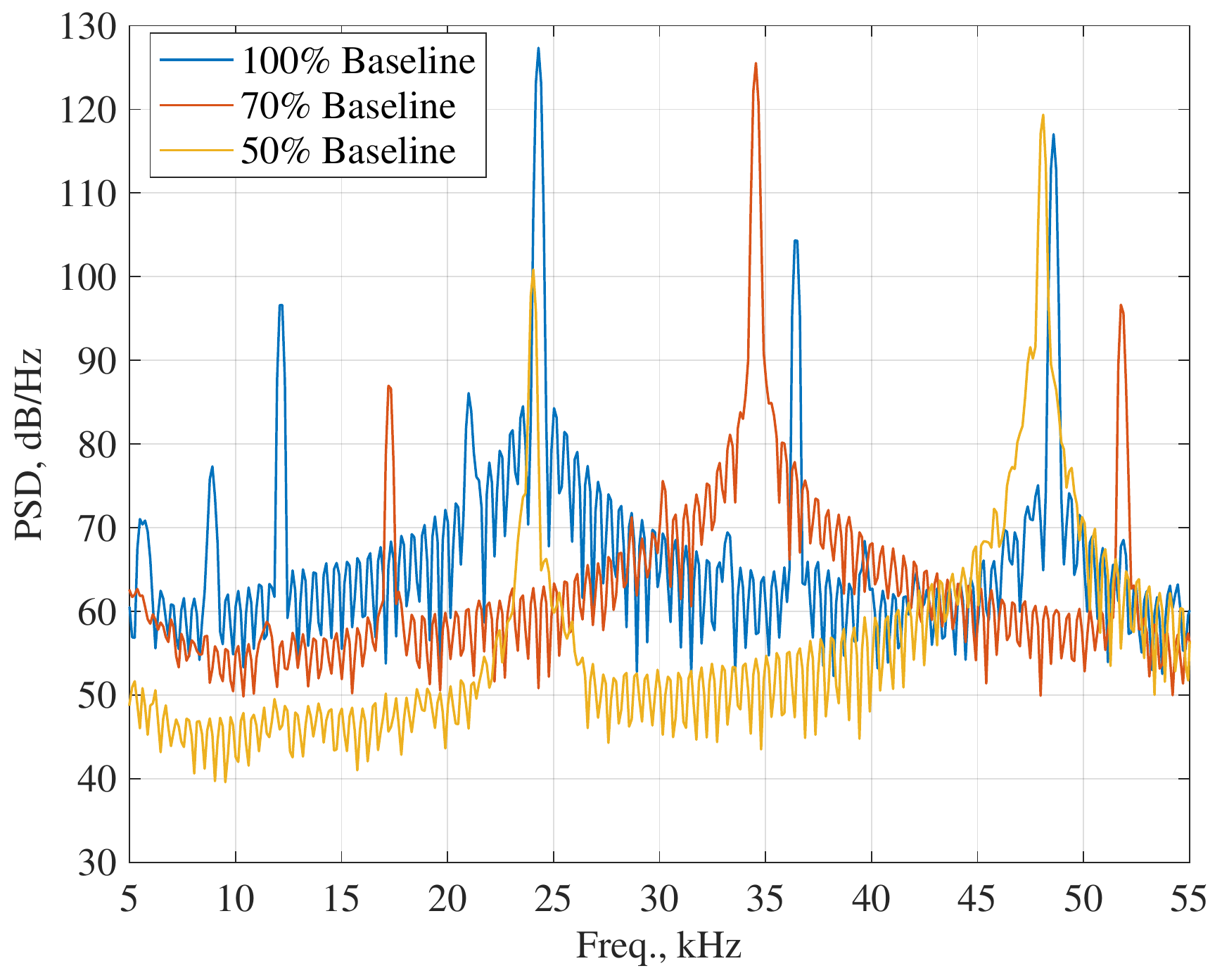}
    \caption{2D CFD predictions of farfield pressure PSD spectra of three whistles derived by scaling the baseline whistle}
    \label{fig:2dCFD_VarSize_3Specs}
\end{figure}

Figure \ref{fig:2dCFD_VarSize_3Specs} shows the farfield pressure PSD spectra of three whistles that are 100\%, 70\%, and 50\% of the baseline whistle. The results show that this set of whistles cover the $20$ kHz to $50$ kHz frequency range, which covers a majority of the bat species in North America adversely impacted by bat-turbine interaction \citep{szewczak2011echolocation}.

\subsubsection*{Three-dimensional aeroacoustic analysis and comparison with data}
\label{sec:3D_CFD}
\begin{figure}[htb!]
    \subcaptionbox{absorbing chamber\label{fig:3DCFD_Mesh_General}}{\incfig[width=0.27\textwidth]{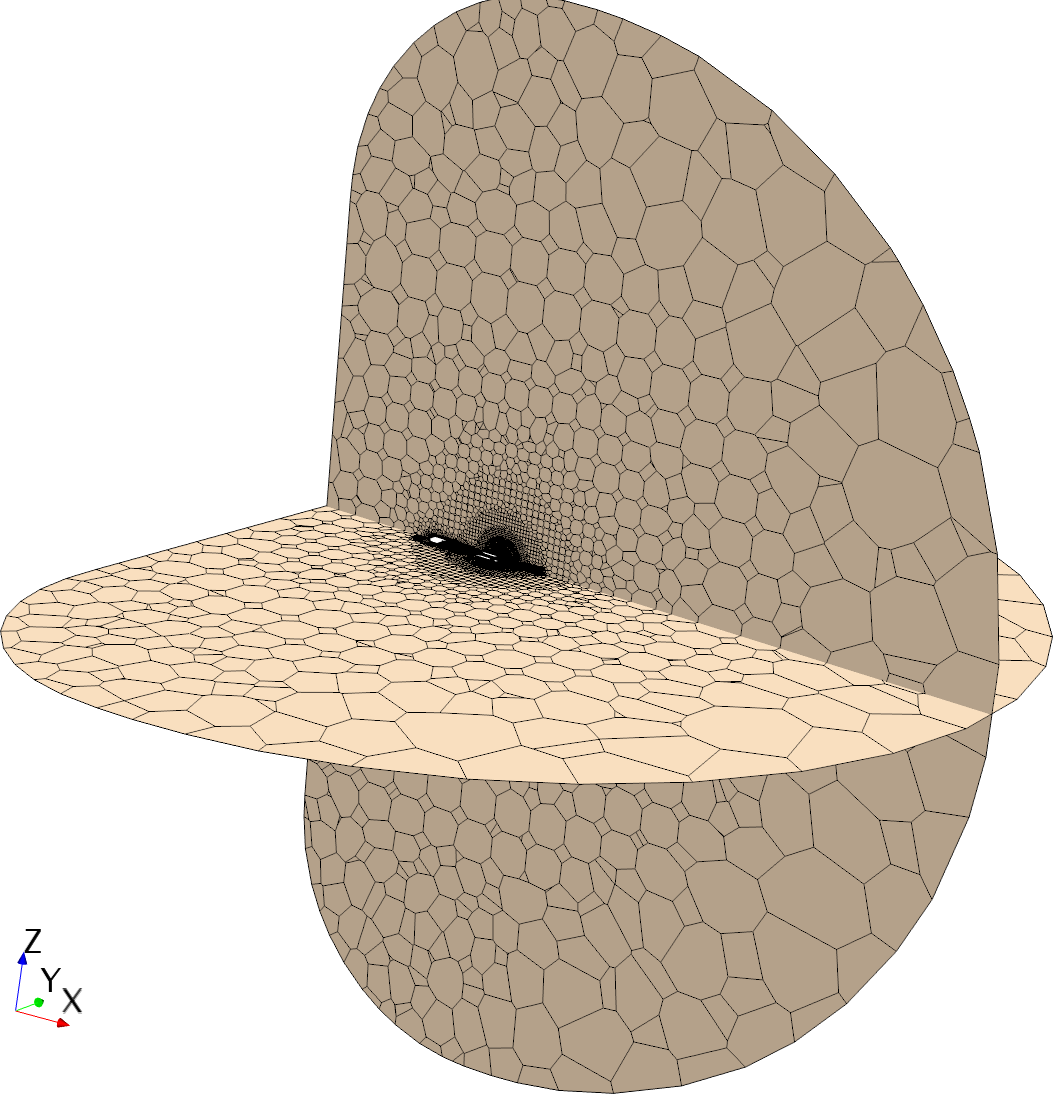}}
    \,
    \subcaptionbox{mesh inside the whistle\label{fig:3DCFD_Mesh_Whistle}}{\incfig[width=0.34\textwidth]{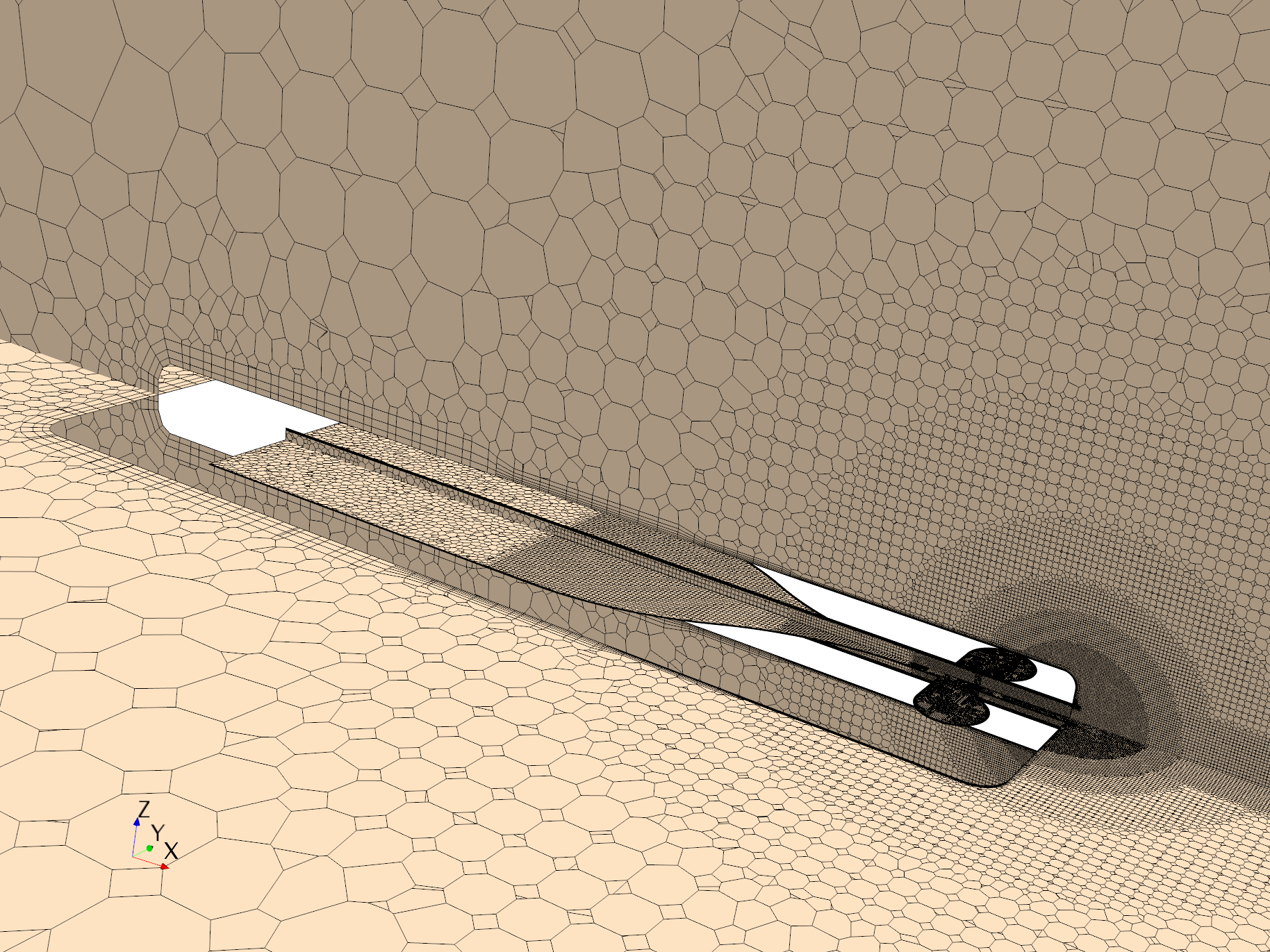}}
    \,
    \subcaptionbox{zoom view of the resonators\label{fig:3DCFD_Mesh_Res}}{\incfig[width=0.34\textwidth]{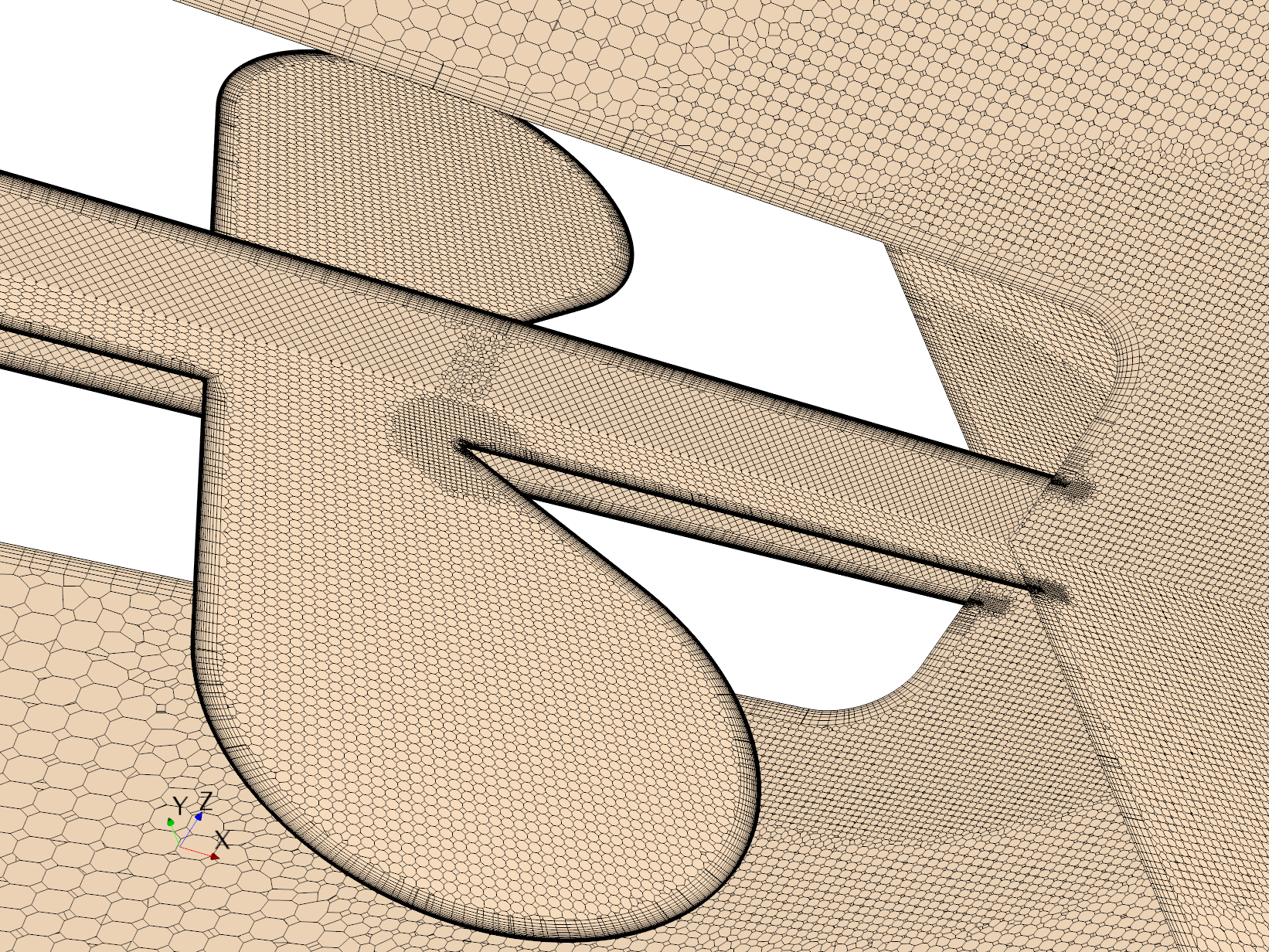}}
    \caption{Computational mesh for 3D CFD analysis of the baseline whistle\label{fig:3DCFD_MeshXY}}
\end{figure}

While the 2D simulations are helpful for qualitative understanding and design guidance, they cannot be quantitatively compared with experimental data. Three-dimensional (3D) simulations are therefore performed to validate the prediction approach. Figure \ref{fig:3DCFD_MeshXY} shows the mesh used for the 3D simulations of the baseline whistle. An absorbing chamber that completely surrounds the whistle is meshed to account for 3D acoustic radiation.

The boundary conditions for the 3D simulations follow the 2D simulations, with total pressure specified at whistle inlet, whistle surfaces treated as adiabatic, no-slip walls, and freestream conditions specified at the outer boundary. A nominal uniform flow speed of $1$ m/s (along the whistle axis) is specified in the absorbing chamber to stabilize the simulation by ensuring that the boundary condition does not fluctuate between inflow and outflow due to small perturbations due to acoustics waves or numerical error. The freestream condition at the outer boundary of the absorbing chamber supports this uniform flow. 

The numerical modeling approach is also the same as in the 2D simulations. The cell size increases rapidly in the absorbing chamber away from the whistle exit (see Fig. \ref{fig:3DCFD_MeshXY}) in order to manage the total mesh size. Excessive numerical dissipation due to coarse grids does not permit accurate acoustic propagation to the farfield in these numerical simulations. Therefore, the farfield acoustic radiation is obtained by solving the FW-H equation (Eq. \ref{eq:FWH}). The FW-H integration surface is shown in Fig. \ref{fig:integration_surface}. The observer locations at which the FW-H predicts the radiated sound are located $1.651$ m from the exit of the whistle, which is the same as the whistle-microphone distance in the measurements. The observers are distributed along a semi-circular array in $15^\circ$ polar angle intervals. 



\begin{figure}[htb!]
    \subcaptionbox{Pressure PSD spectra
    \label{fig:3DCFD_SpecCompare_PSD}}{\incfig[width=0.49\textwidth]{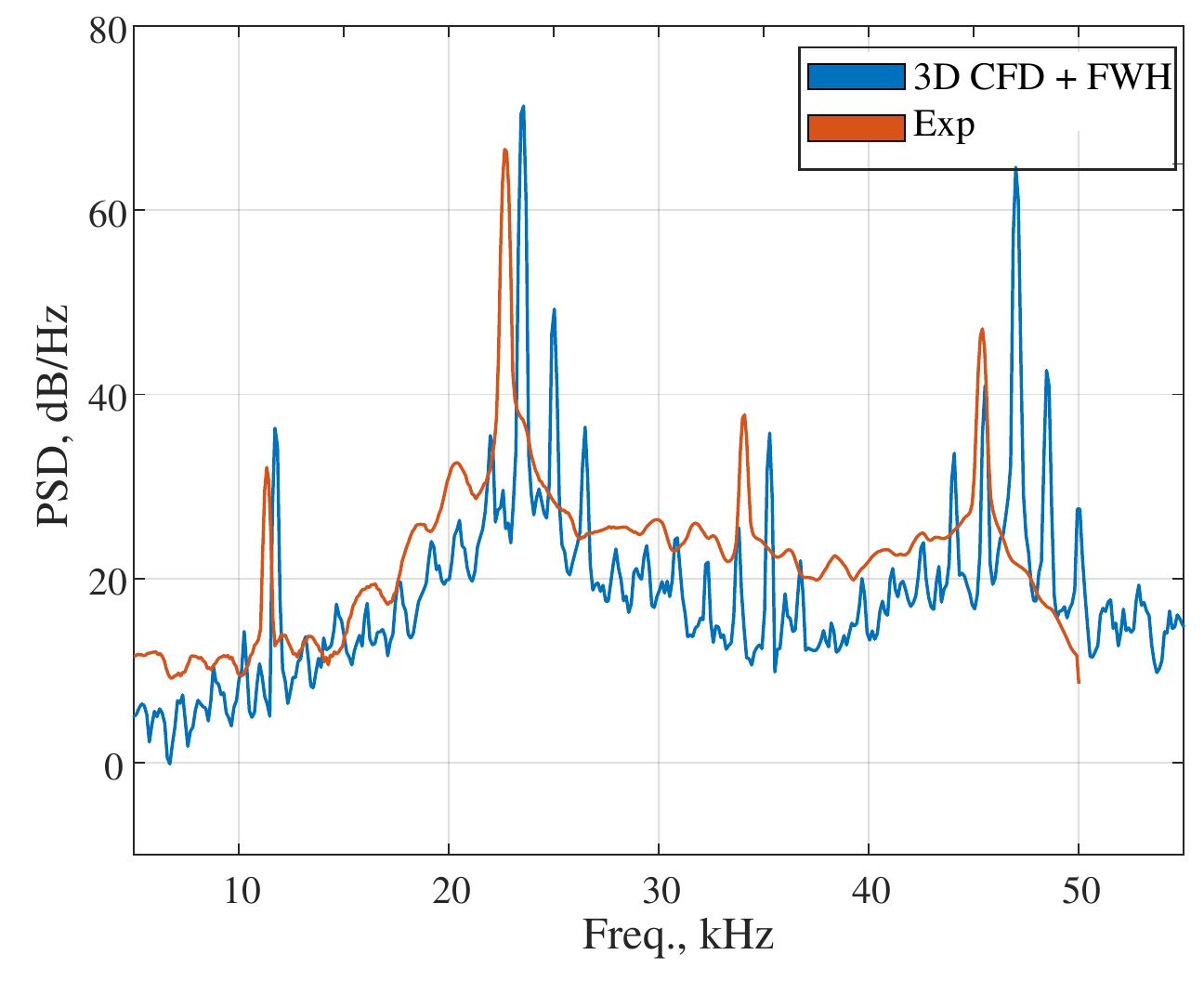}}
    \hfill
    \subcaptionbox{SPL of dominant tones
    \label{fig:3DCFD_SpecCompare_SPL}}{\incfig[width=0.49\textwidth]{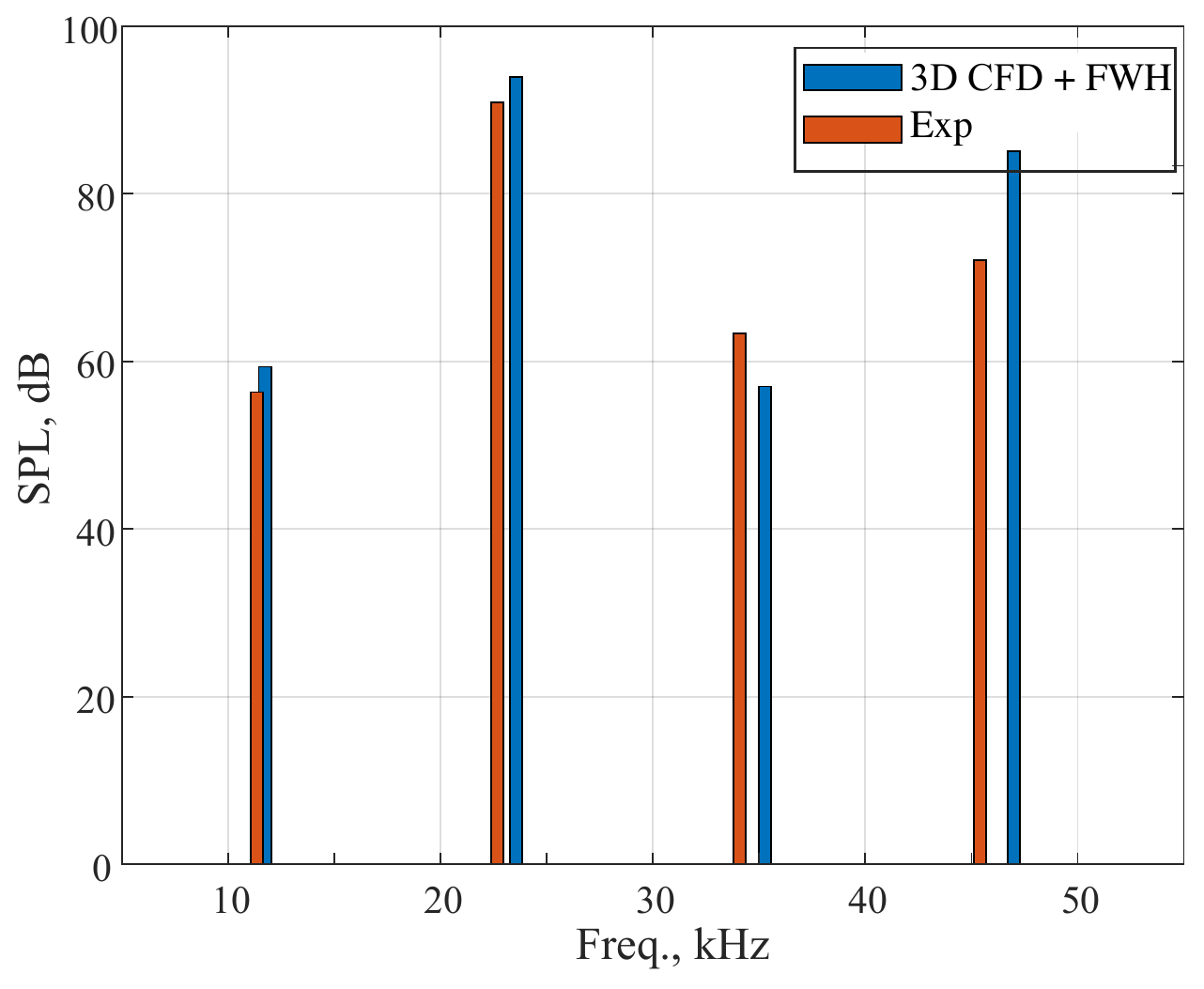}}
    \caption{Comparison of the radiating ultrasound between measurements in the anechoic chamber and numerical (3D CFD + FW-H) results: (a) pressure PSD spectra in the farfield at $30^\circ$ polar angle, and (b) SPL of dominant tones\label{fig:3DCFD_SpecCompare}}
\end{figure}

Figure \ref{fig:3DCFD_SpecCompare} compares the numerical acoustic predictions (using 3D CFD and FW-H) with the measured spectra at $30$ deg polar angle and $1.651$ m away from the whistle. Pressure PSD and tonal SPL spectra are compared in the figure. The predictions agree reasonably well with the measurements, particularly for the tonal SPL spectra. The simulations show a slightly higher peak frequency than the measurements, which could be due to differences in modeled and fabricated geometries of the whistles, and in the supply pressure. Nevertheless, the peak amplitudes are quite close. The larger scatter in the simulations is because the time over which the simulated spectra is averaged is not as large as for the measured spectra.

\begin{figure}[htb!]
    \subcaptionbox{directivity of the dominant peak\label{fig:3DCFD_PeakPolar}}{\incfig[width=0.5\textwidth]{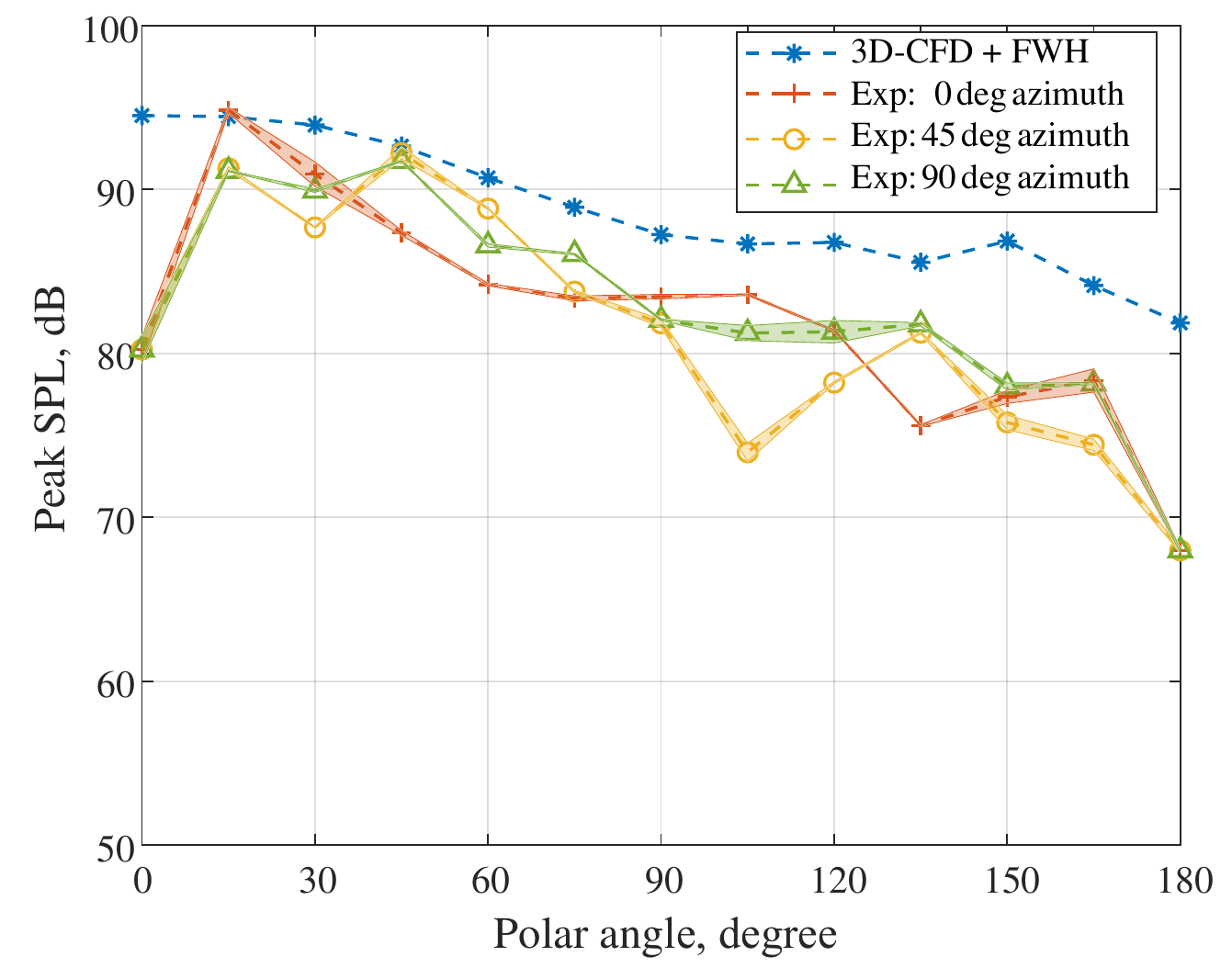}}
    \hfill
    \subcaptionbox{radiated acoustic power PSD\label{fig:3DCFD_SWL_FWHvsExp}}{\incfig[width=0.5\textwidth]{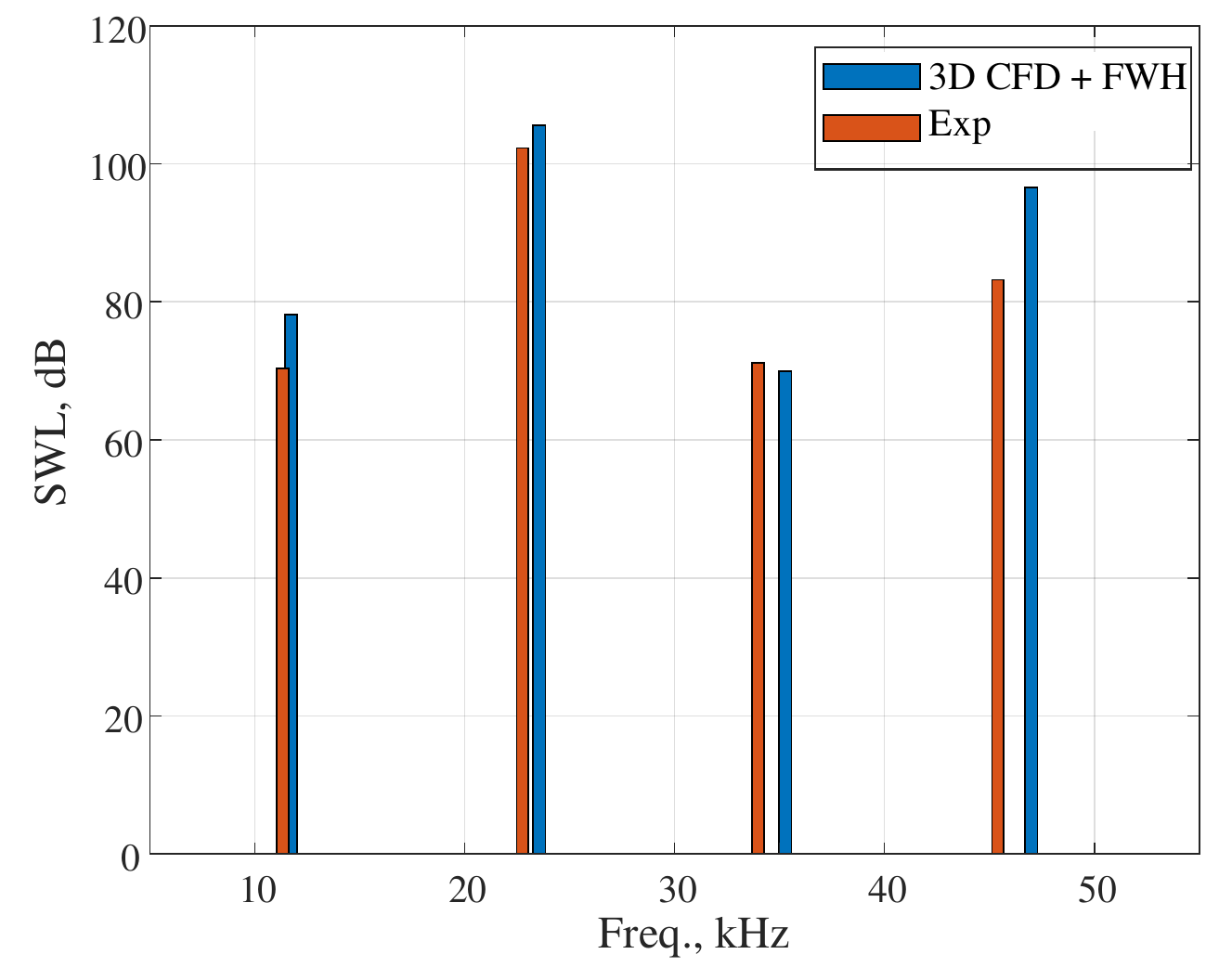}}
    \caption{Comparisons between 3D-CFD FWH and measurements (a) peak SPL polar directivity (at 0$^\circ$ azimuthal angle, $1.651$ m from whistle exit), and (b) sound power level (SWL) spectra (wref E-12 W).
    \label{fig:PWL_3DCFD_vs_Exp}}
\end{figure}

Figure~\ref{fig:3DCFD_PeakPolar} compares the farfield polar directivity of peak SPL between simulations and experiments for a fixed azimuthal angle ($=0^\circ$). The predictions qualitatively capture the measured behavior -- SPL decreases with increasing polar angle. However, the sharp reduction observed at $0^\circ$ and $180^\circ$ in the measurements is not seen in the predictions. This sharp reduction in radiated ultrasound along the axis in the experiments is a result of acoustic refraction due to whistle exhaust flow at $0^\circ$, and due to scattering/blockage by the whistle and the stand (to hold the whistle in place) at $180^\circ$. These refraction and scattering effects are not included in the FW-H acoustic radiation model.

The power in the radiated ultrasound field is obtained by integrating the flux of acoustic power through a spherical surface surrounding the whistle. Figure \ref{fig:3DCFD_SWL_FWHvsExp} compares the radiated acoustic power PSD between the predictions and experiments.
Atmospheric attenuation is accounted for in the simulations. The predicted radiated ultrasound power levels (SWL) are slightly higher than the measurements.
%

\section*{Six-whistle ultrasound deterrent}
\label{sec:6whistles}
\begin{figure}[htb!]
    \centering
    \subcaptionbox{CAD model\label{fig:6WS_AssmGeneral}}{\incfig[width=0.4\textwidth]{Figs/6WhsDet_CADAssem.jpg}}
    \qquad
    \subcaptionbox{3D-printed specimen\label{fig:6WS_Device}}{\incfig[width=0.35\textwidth]{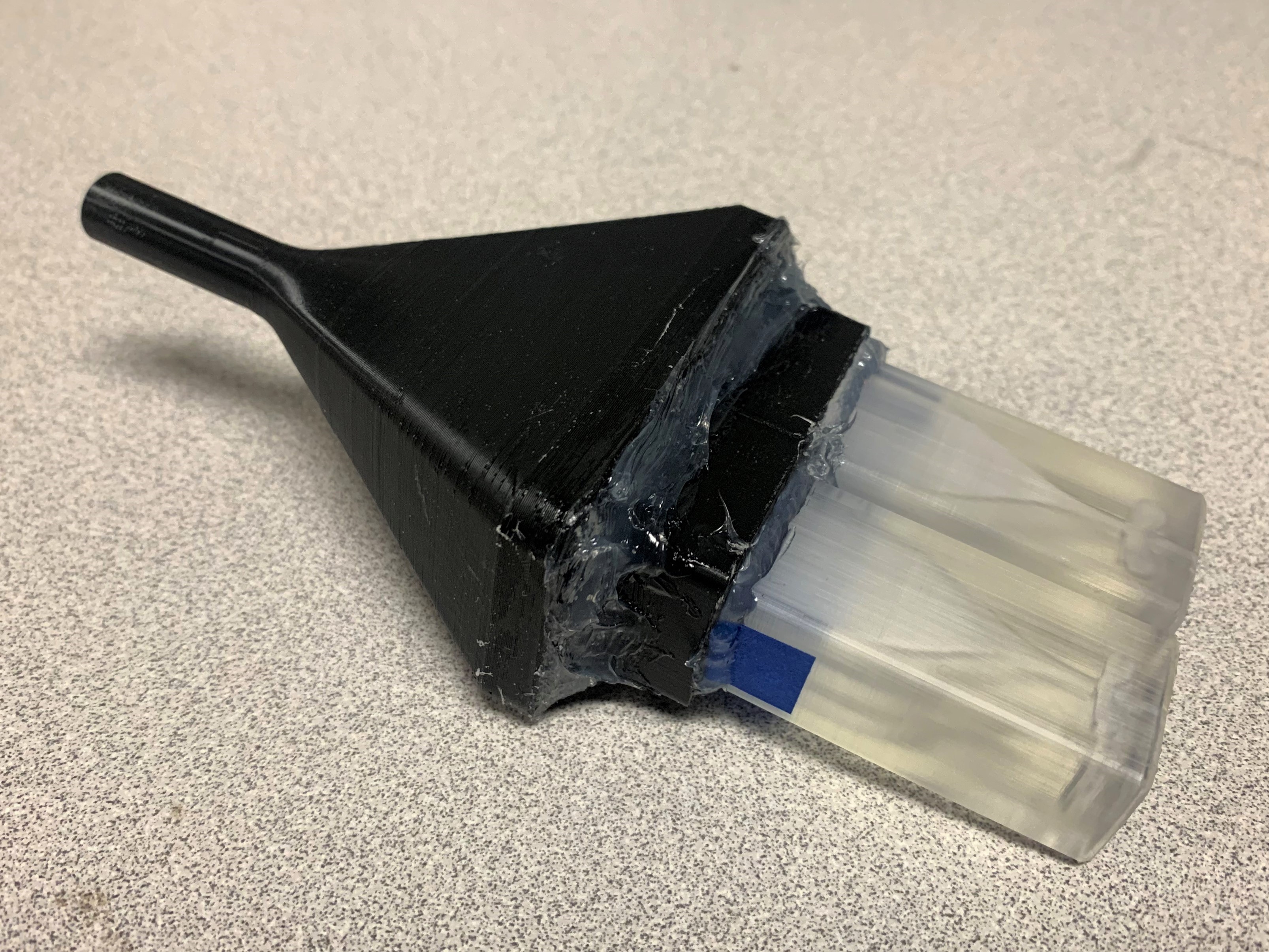}}
    \caption{A six-whistle active ultrasonic deterrent device. (a) a computer-aided design (CAD) model of the deterrent, and (b) the model fabricated using additive manufacturing (3D printing). \label{fig:6WS_Assm}}
\end{figure}

An ultrasonic bat deterrent targeting multiple frequencies in the $20-50$ kHz range is developed based on the baseline whistle. It consists of six whistles that are obtained by geometrically scaling the baseline design. The whistles are scaled 100\%, 85\%, 70\%, 60\%, 55\% and 50\% of the baseline along each dimension. All the whistles are powered by one source of pressurized air. Since the throat area of the whistle determines the flow rate through it, the depth of the whistles are modified to ensure identical throat areas ($3.2~\rm mm^{2}$) for the six whistles. For ease of fabrication, the six whistles are distributed into two sets, with each set containing three whistles (see Fig. \ref{fig:6WS_Assm}). The two whistle sets are connected to a diffuser via a socket and pressurized air is supplied to the whistles through the diffuser. Each whistle set is 3D printed as one piece using clear resin. The diffuser and socket are 3D-printed using polylactic acid (PLA). Figure \ref{fig:6WS_Device} is a picture of the fabricated deterrent. The different parts of the deterrent are assembled and held together using epoxy. 

A significant advantage of this deterrent is its low cost. As we have done here, the deterrent can be fabricated with 3D printing using plastic (PLA) instead of machined metal, and has a very simple design/structure. Therefore the production cost is much lower than that of transducer-based ultrasonic deterrents. Secondly, the proposed deterrent is readily driven by off-the-shelf air compressors; there is no need for delicate and fragile electronic transducers/controllers. The operational cost of this deterrent is therefore also expected to be lesser.
 
\subsection*{Experimental analysis of the six-whistle deterrent}
\label{sec:Exp_6WS}
\begin{figure}[htb!]
    \subcaptionbox{PSD spectra\label{fig:6WS_3Specs}}{\incfig[width=0.49\textwidth]{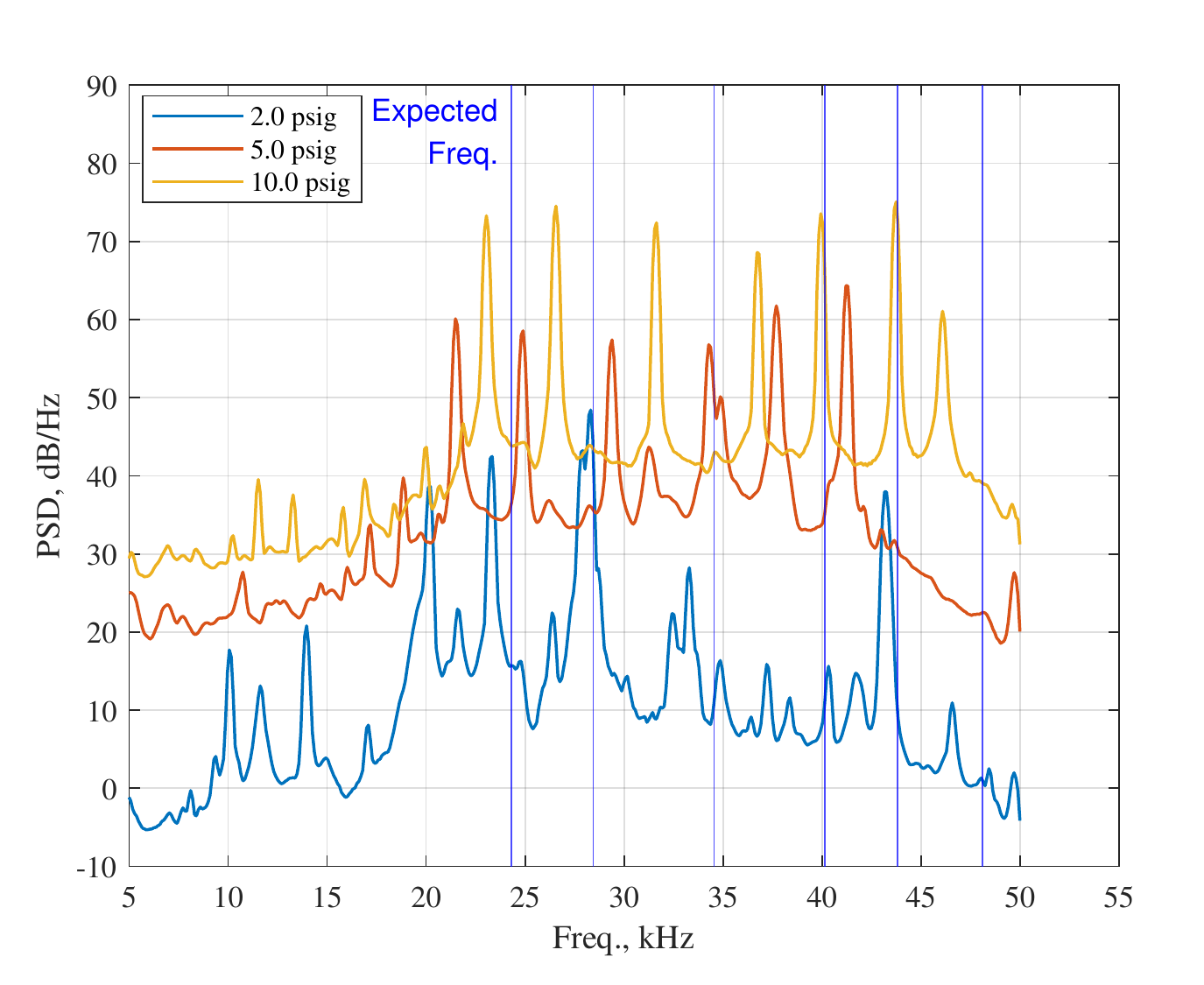}}
    \hfill
    \subcaptionbox{SPL spectra\label{fig:6WS_3SPLHist}}{\incfig[width=0.49\textwidth]{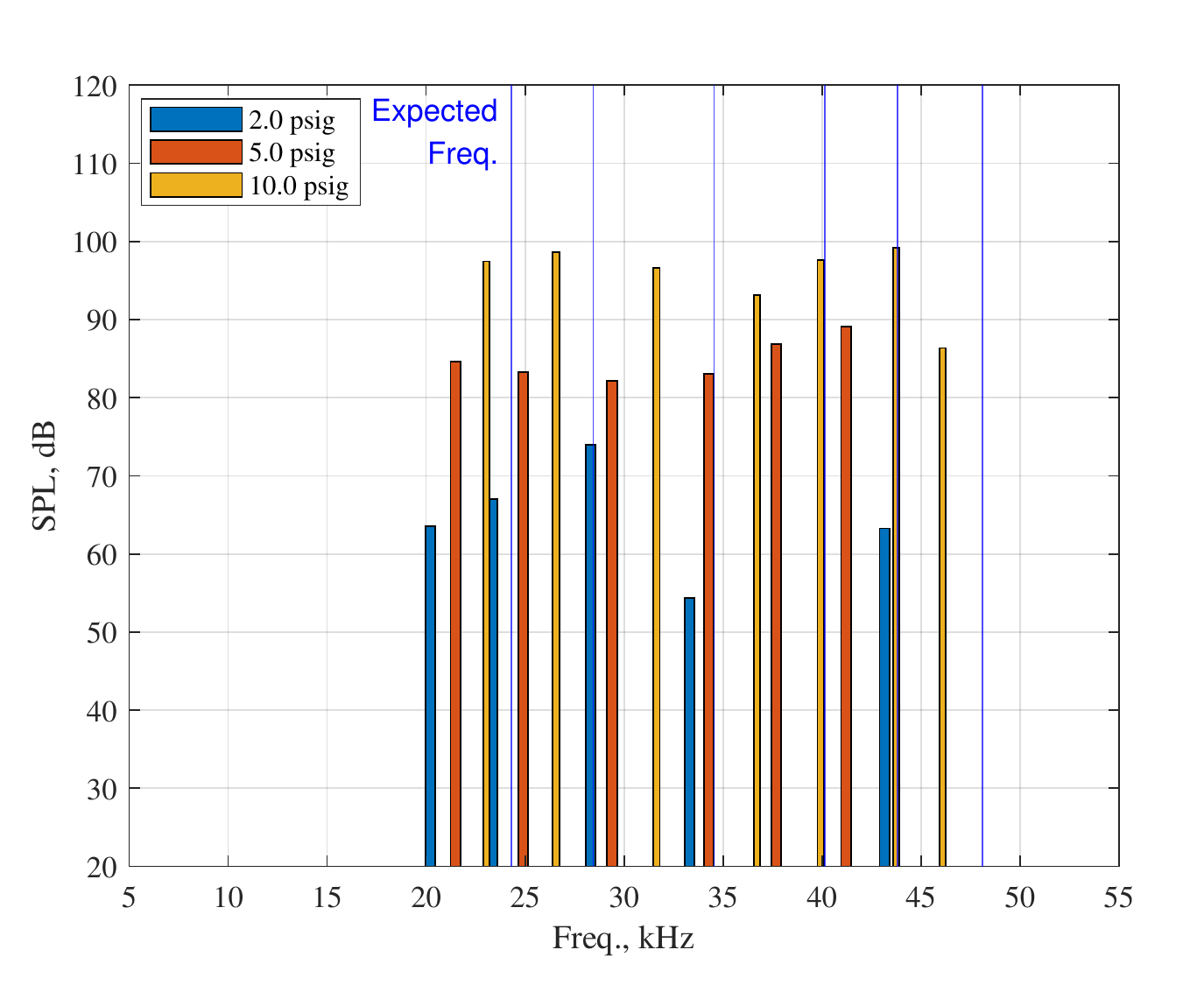}}
    \caption{Experimental results of the six-whistle active ultrasonic deterrent for three supply air pressures. (a) power spectral density (PSD) spectra, and (b) sound pressure level (SPL) spectra. Expected frequencies from isolated-whistle analysis are shown as vertical blue lines.\label{fig:6WL_exp_results}}
\end{figure}

The six-whistle deterrent is tested in the anechoic chamber using the same setup as for the baseline whistle. The deterrent is expected to generate at least six major peak frequencies (see Table \ref{tab:VariousSize}). The acoustic performance of this deterrent is tested for supply pressure ($p_t$) of $2,~3,~4,~5,~7$ and $10$ psig. The microphone capturing the radiating sound is located at $30^\circ$ polar angle. 
Figure \ref{fig:6WS_3Specs} plots the farfield pressure PSD spectra for $p_t=2,~5$ and $10$ psig. It is seen that although a six-peak-pattern in the spectrum is established for $p_t\geq 5$ psig, the frequencies of the peaks (even at $10$ psig), are slightly lower than the values expected from isolated-whistle analysis.

Figure~\ref{fig:6WS_3SPLHist} shows the SPLs of the significant peaks for three values of $p_t$ (=$2,5,10$ psig). At $10$ psig, the six major peaks reach SPL values between $90-100$ dB, which is significantly higher than those at $5$ psig. 
With higher supply pressure ($50$ psig possible with off-the-shelf air compressors), it is possible to match the ultrasound radiation levels (SPL $\sim 115$ dB) of commercially available electromechanical transducer based ultrasonic deterrents.

\section*{Conclusion}
\label{sec:conclusion}
Numerical and experimental analyses of novel ultrasonic deterrents designed using aerodynamic whistles are presented. The baseline deterrent/whistle, which consists of two identical resonators, generates ultrasound with a peak frequency around $24$ kHz when driven by pressurized air. The supply air pressure ($p_t$) is varied between $0.5-2.0$ psig. Farfield acoustic measurements in an anechoic chamber show that the peak frequency increases very slightly with $p_t$. Two-dimensional CFD simulations of the baseline deterrent reveal that, in the $p_t$ range explored here, the two resonating chambers of the whistle operate out-of-phase and the peak frequency of pressure fluctuations in each resonator is about $12$ kHz, which is half that of the farfield radiating ultrasound. This flow pattern, as well as the peak frequency, are found to be insensitive to variations in $p_t$. These findings suggest that the mechanism of ultrasound generation in the baseline deterrent is Helmholtz resonance.

The numerical prediction approach is validated with farfield acoustic measurements in the anechoic chamber. Three-dimensional CFD simulation results are used with the Ffowcs Williams-Hawkings acoustic analogy to predict farfield ultrasound radiation. Pressure power spectral density spectra, tonal sound pressure level (SPL) spectra, and polar directivity are compared. The tonal SPL levels agree well although there are some discrepancies in directivity.  

A six-whistle ultrasound deterrent is designed by geometrically scaling the baseline whistle. The whistles are 100\%, 85\%, 70\%, 60\%, 55\% and 50\% the size of the baseline whistle.  The deterrent is fabricated and the model is tested in the anechoic chamber over the supply air pressure range $2.0 \le p_t \le 10.0$ psig. As $p_t$ is increased above $5.0$ psig, six dominant peaks are observed in the farfield ultrasound in the $20-50$ kHz range. The SPL values of the peaks reach $90$ dB with a reference pressure of $20 \mu$pa at a distance $1.651$ m away from the whistle when $p_t=10$ psig. 
%

\section*{Acknowledgments}
\label{sec:acknowledgments}
Funding for this research is provided by the National Science Foundation (Grants CBET-1554196 and 1935255). Computational resources are provided by NSF XSEDE (Grant \#TG-CTS130004) and the Argonne Leadership Computing Facility, which is a DOE Office of Science User Facility supported under Contract DE-AC02-06CH11357.
\bibliographystyle{elsarticle-num-names} 
\bibliography{Refs}

\pagebreak
\appendix
\section*{Appendices}
\label{app:appendices}
\subsection*{Appendix A: Theoretical estimates of frequencies in classical resonators}
\label{app:ResFreq}
\begin{figure}[htb!]
    \incfig[width=.2\textwidth]{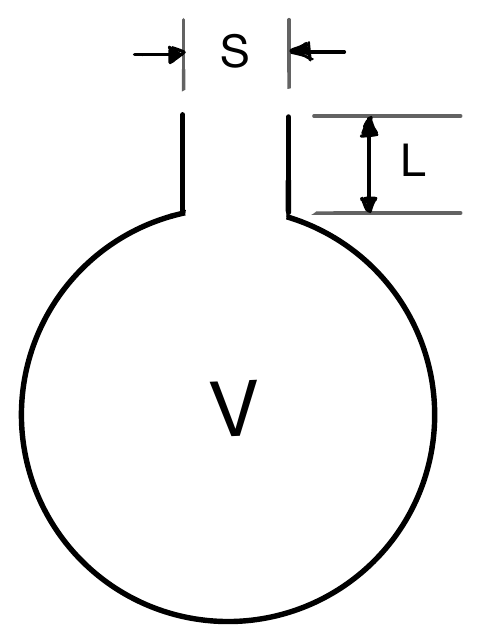}
    \caption{A schematic of a Helmholtz resonator\label{fig:HelmRes}}
\end{figure}

We provide theoretical estimates of frequencies corresponding to Helmholtz resonance, standing-wave cavity resonance, Rossiter modes \cite{rossiter1964wind} and pipe resonance mechanisms. A typical Helmholtz resonator \cite{kinsler1999fundamentals} consists of a rigid-walled cavity of volume $V$ neck area $S$ and neck length $L$ (Fig. \ref{fig:HelmRes}).
A Helmholtz resonator can be treated as a spring-mass dynamical system where the air in the neck serves as the system ``mass'' and the compressibility of the fluid (air) in the cavity plays the role of the ``spring''. The natural frequency of this system, $f_{\rm HR}$ can be expressed as
\begin{equation}
    f_{\rm HR} = \frac{c}{2 \pi}\sqrt{\frac{S}{V L_s}},
    \label{eqn: Helmholtz Freq}
\end{equation}
where $c$ is the speed of sound, $L_s$ is the effective neck length. In the typical model $L_s = L$, but in practical situations $L_s$ is obtained using $L$ and a linear scale of the neck area~\cite{kinsler1999fundamentals,ruijgrok1993elements}. For resonators in which the neck is formed of a thin wall \ie $L$ is very small, \citet{ma2009fluid} suggests using an ``effective'' neck length equal to the opening length of the neck to calculate $f_{\rm HR}$. Taking $L_s = O$ for the resonators in the baseline whistle (see Fig. \ref{fig:Baseline_2DdesignA}), the $f_{\rm HR}$ is estimated to be $11.7$ kHz.

The cavity closed by the shear layer can respond according to its box/cavity modes, which are standing waves in the longitudinal, crossflow (along the depth), and spanwise directions \cite{gloerfelt2009cavity}. In our case, the depth-mode standing wave frequency, $f_{\rm SW}$, is estimated using
\begin{equation}
    f_{\rm SW} = \frac{c}{2}\,\frac{n_{\rm depth}}{2 (L_{\rm chamber} - t/2)},
    \label{eq:standing_wave}
\end{equation}
where $n_{depth}$ is an integer. When $n_{depth} = 1$, the estimated $f_{SW}$ for the resonators of the baseline whistle is $15.3$ kHz. Note that this equation assumes a rectangular cavity.

The frequencies of the Rossiter modes, $f_{\rm RM}$ are estimated using 
\begin{equation}
    f_{\rm RM} = \frac{U}{O}\,\frac{n-\alpha}{(M + 1/\kappa)},
    \label{eq:rossiter_mode}
\end{equation}
where $\alpha$ describes the phase delay between the hydrodynamic forcing and the acoustic feedback and $\kappa$ is the convection velocity of the shear layer normalized by the free stream velocity and $n$ is an integer \cite{bennett2017resonant}. Since the flow Mach number ($M$) is small in our application, $\alpha = 0$. Empirical value of $\kappa = 0.5$ is used, and the flow speed ($U$) and $M$ are obtained from CFD data acquired at the center of the whistle throat. 

\subsection*{Appendix B: Rossiter modes}
\label{app:rossiter}
\begin{figure}[htb!]
    \subcaptionbox{Design\label{fig:Ross_design}}{\incfig[width=0.25\textwidth]{Figs/ModLargeDesign_2D.PNG}}
    \hfill
    \subcaptionbox{Peak freq. variation with $p_t$\label{fig:Ross_peakFreq}}{\incfig[width=0.74\textwidth]{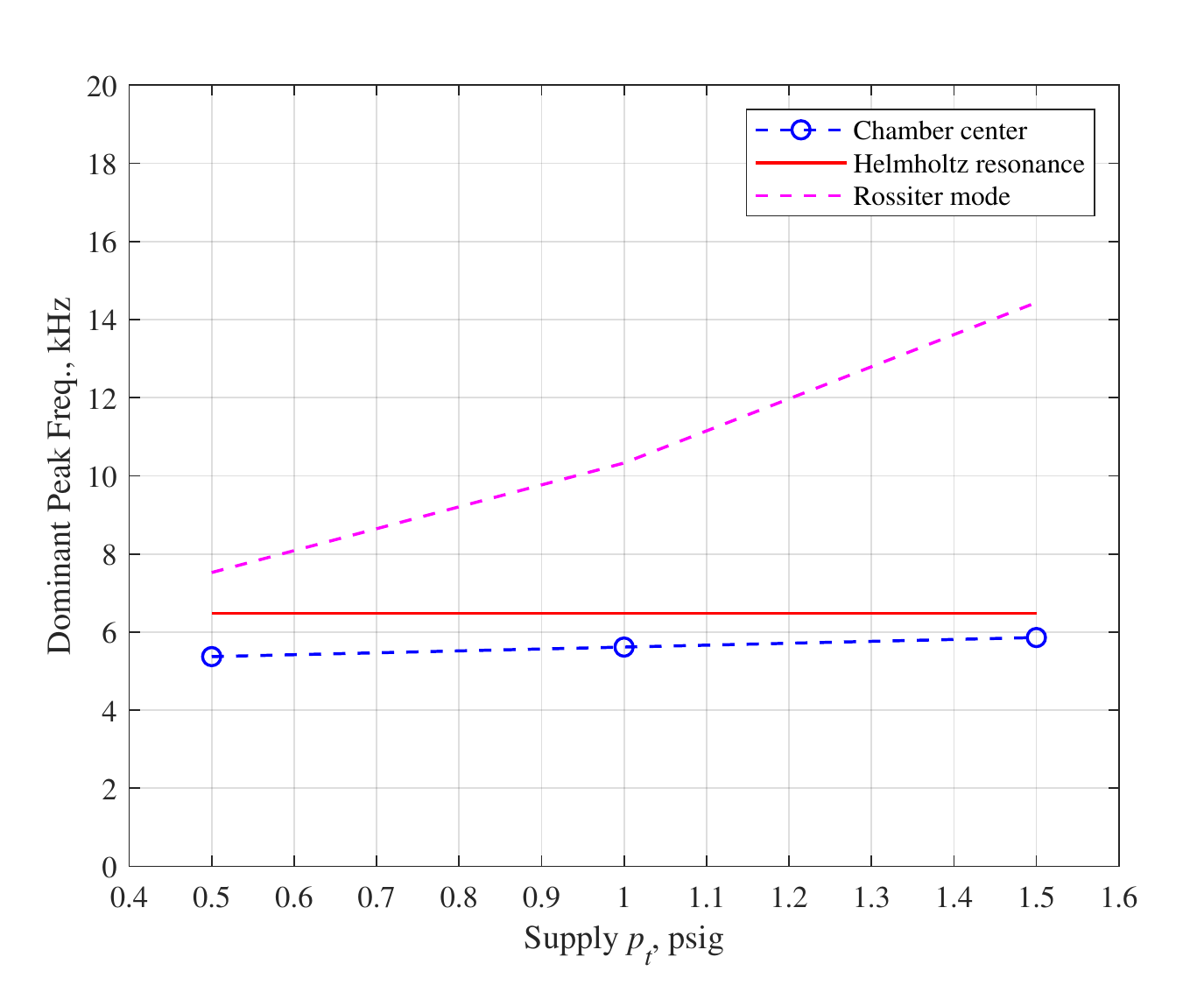}}
    \caption{
    \label{fig:HRvsRoss}}
\end{figure}

For the baseline design evaluated in this paper, the predicted first Rossiter's mode frequency coincides with the Helmholtz resonance frequency of the chamber in the range of supply pressure ($p_t$) tested. To separate these two frequencies, a whistle with a modified geometry (Fig. \ref{fig:Ross_design}) is simulated for $p_t = 0.5, 1.0, 1.5$ psig.  The modified whistle chamber has a larger volume compared with the baseline design, reducing the Helmholtz resonance frequency to $0.65$ kHz; the length of the gap over the orifice ($O$) and the angle of the downstream edge ($\theta$) edge are unchanged.

Figure \ref{fig:Ross_peakFreq} shows the variation with $p_t$ of the predicted peak frequency at the center of the chamber. The expected frequencies corresponding to the first Rossiter mode and Helmholtz resonance, are also plotted. The predicted chamber frequency closely follows the Helmholtz resonance frequency, indicating that Helmholtz resonance is the dominant mechanism here.
\bibliographystyle{plos2015} 
\bibliography{Refs}
\end{document}